\newif\ifarxiv
\arxivtrue

\ifarxiv
\documentclass[a4paper, cleveref, autoref, thm-restate]{lipics-v2021}
\pdfoutput=1
\hideLIPIcs
\else
\documentclass[runningheads]{llncs}
\fi

\usepackage{tikz}
\usetikzlibrary{external}
\ifarxiv
\else
\fi

\usepackage[T1]{fontenc}

\usepackage{rotating}
\usepackage{cite}
\usepackage{amsmath,amssymb,amsfonts}
\usepackage{algorithmic}
\usepackage{graphicx}
\usepackage{textcomp}
\usepackage{xcolor}
\usepackage{mathtools}
\usepackage{hyperref}
\usepackage{cleveref}
\usepackage{siunitx}
\usepackage{numprint}
\npdecimalsign{.}
\npthousandsep{\,}

\usepackage[algo2e, ruled, vlined, linesnumbered]{algorithm2e}
\usepackage{booktabs}
\usepackage[alpine, misc, clock]{ifsym}
\usepackage{float}
\usepackage{multirow}
\usepackage{wasysym}

\def\etal.{et\penalty50\ al.}
  
\newif\ifdoubleblind

\newcommand{\partitioner}[1]{\textsf{#1}}
\newcommand{\algname}{\partitioner{dKaMinPar}}
\newcommand{\algnameFast}{\partitioner{dKaMinPar-Fast}}
\newcommand{\algnameStrong}{\partitioner{dKaMinPar-Strong}}

\newcommand{\R}[1]{\textcolor{blue}{#1}}
\renewcommand{\R}[1]{#1}

\newcommand{\symbTimeout}{\clock}
\newcommand{\symbInfeasible}{\ensuremath{\times}}

\newcommand{\rggTwoD}[2]{\ensuremath{\textsf{rgg}_{\textrm{2D}}#1\textrm{d}#2}}
\newcommand{\rggThreeD}[2]{\ensuremath{\textsf{rgg}_{\textrm{3D}}#1\textrm{d}#2}}
\newcommand{\rhg}[3]{\ensuremath{\textsf{rhg}_{#3}#1\textrm{d}#2}}

\newcommand{\tblSymbTimeout}{\clock}
\newcommand{\tblSymbInfeasible}{\ensuremath{\triangle}}
\newcommand{\tblSymbFailed}{\ensuremath{\times}}

\newcommand{\dkaminparFastFailsLargek}{0}
\newcommand{\dkaminparFastInfeasiblesLargek}{0}
\newcommand{\dkaminparFastTimeoutsLargek}{0}
\newcommand{\dkaminparFastFeasiblesLargek}{128}
\newcommand{\kaminparFailsLargek}{0}
\newcommand{\kaminparInfeasiblesLargek}{0}
\newcommand{\kaminparTimeoutsLargek}{0}
\newcommand{\kaminparFeasiblesLargek}{128}
\newcommand{\parhipFastFailsLargek}{48}
\newcommand{\parhipFastInfeasiblesLargek}{57}
\newcommand{\parhipFastTimeoutsLargek}{6}
\newcommand{\parhipFastFeasiblesLargek}{17}
\newcommand{\parhipEcoFailsLargek}{40}
\newcommand{\parhipEcoInfeasiblesLargek}{37}
\newcommand{\parhipEcoTimeoutsLargek}{39}
\newcommand{\parhipEcoFeasiblesLargek}{12}
\newcommand{\parmetisFailsLargek}{44}
\newcommand{\parmetisInfeasiblesLargek}{54}
\newcommand{\parmetisTimeoutsLargek}{0}
\newcommand{\parmetisFeasiblesLargek}{30}
\newcommand{\kaminparRunningTimeLargek}{0.46}
\newcommand{\parhipFastRunningTimeLargek}{22.21}
\newcommand{\parhipEcoRunningTimeLargek}{88.35}
\newcommand{\parmetisRunningTimeLargek}{2.38}
\newcommand{\kaminparCutLargek}{0.98}
\newcommand{\parhipFastCutLargek}{1.06}
\newcommand{\parhipEcoCutLargek}{0.96}
\newcommand{\parmetisCutLargek}{1.01}
\newcommand{\parhipFastGmeanInfImbalanceLargek}{1.19}
\newcommand{\parhipEcoGmeanInfImbalanceLargek}{1.10}
\newcommand{\parmetisGmeanInfImbalanceLargek}{1.18}

\newcommand{\smallkInstances}{224}

\newcommand{\parmetisSmallkInfeasiblesPercentage}{34}
\newcommand{\smallkCommonRunningTimes}{145}
\newcommand{\smallkCommonRunningTimedKaMinParFast}{4.93}

\newcommand{\smallkCommonRunningTimeParHIPFast}{16.77}

\newcommand{\smallkCommonRunningTimeParMETIS}{6.98}
\newcommand{\smallkCommonRunningTimeParMETISTodKaMinParFast}{1.4}
\newcommand{\smallkCommonRunningTimeParHIPFastTodKaMinParFast}{3.4}

\newcommand{\ssdKaMinParGmeanDecline}{2.0}

\newcommand{\wsLargekdKaMinParToParMETISPercent}{19.3}
\newcommand{\wsLargekdKaMinParToParHIPPercent}{2.8}

\ifarxiv
\title{Distributed Deep Multilevel Graph Partitioning}
\fi

\ifarxiv
\newcommand{\myparagraph}[1]{\subparagraph*{#1}}
\else
\newcommand{\myparagraph}[1]{\paragraph{#1}}
\fi

\ifarxiv 
\author{Peter {Sanders}}{Karlsruhe Institute of Technology, Germany}{sanders@kit.edu}{}{}
\author{Daniel {Seemaier}}{Karlsruhe Institute of Technology, Germany}{daniel.seemaier@kit.edu}{}{}
\authorrunning{P. Sanders, D. Seemaier}

\supplement{The source code and data has been made available at \url{https://algo2.iti.kit.edu/seemaier/ddeep_mgp/}.}

\acknowledgements{This work was performed on the HoreKa supercomputer funded by the Ministry of Science, Research and the Arts Baden-Württemberg and by the Federal Ministry of Education and Research. 
This project has received funding from the European Research Council (ERC) under the European Union’s Horizon 2020 research and innovation programme (grant agreement No. 882500).
\\ \hspace*{\fill} \includegraphics[width = 3cm, trim = {0px 5cm 0px 5cm}, clip]{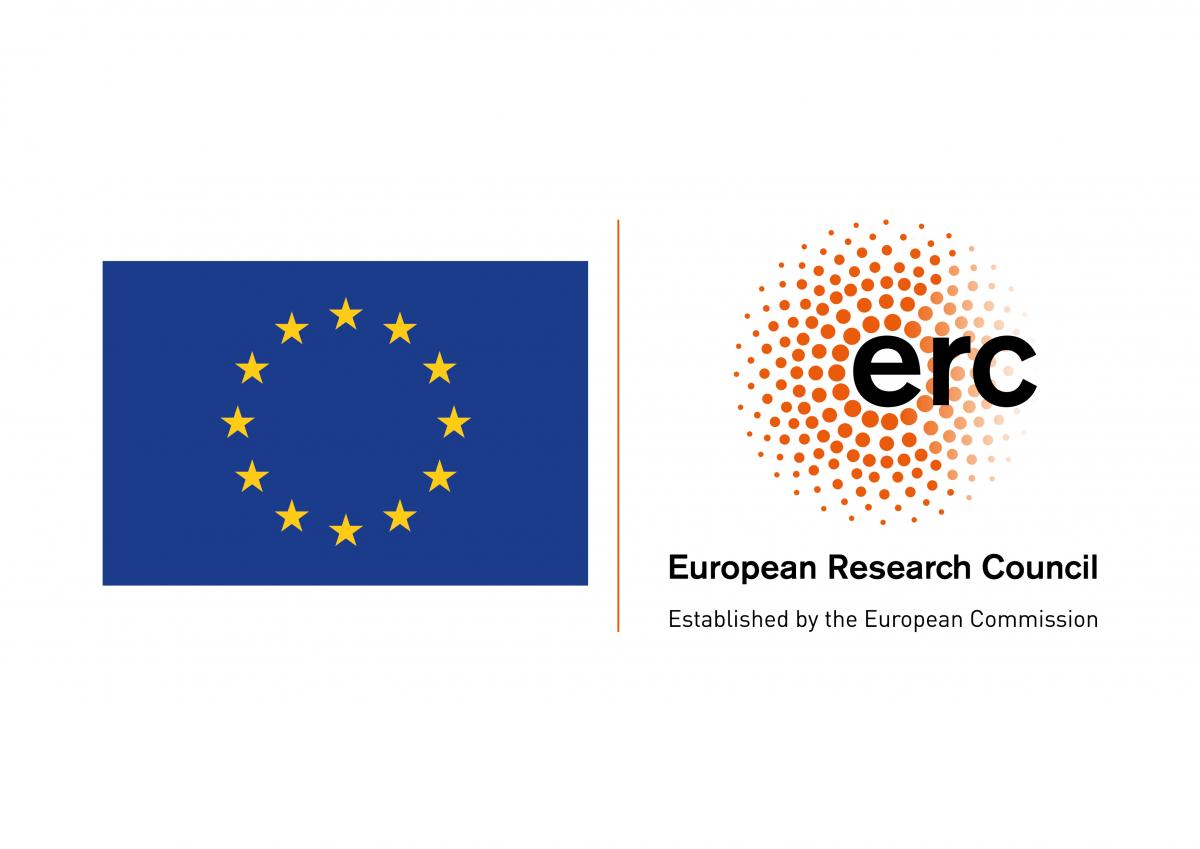} \textcolor{white}{.}
\vspace{-1cm}
}


\nolinenumbers
\fi

\begin{document}

\maketitle

\begin{abstract}
We describe the engineering of the distributed-memory multilevel graph partitioner \algname. 
It scales to (at least) 8192 cores while achieving partitioning quality comparable to widely used sequential and shared-memory graph partitioners. 
In comparison, previous distributed graph partitioners scale only in more restricted scenarios and often induce a considerable quality penalty compared to non-distributed partitioners.
When partitioning into a large number of blocks, they even produce infeasible solution that violate the balancing constraint.
\algname\ achieves its robustness by a scalable distributed implementation of the deep-multilevel scheme for graph partitioning. 
Crucially, this includes new algorithms for balancing during refinement \emph{and} coarsening.

\end{abstract}

\section{Introduction}

Graphs are a central concept of computer science used whenever we need to model relations between objects.
Consequently, handling \emph{large} graphs is very important for parallel processing.
This often requires to \emph{partition} these graphs into blocks of approximately equal weight with most edges inside the blocks (balanced graph partitioning).
Applications include scientific computing, handling social networks, route planning, and graph databases~\cite{RecentAdvances}.

In principle, \emph{multilevel graph partitioners} (MGP) achieve high quality partitions for a wide range of input graphs $G$ with a good trade-off between quality and partitioning cost.
They are based on first iteratively \emph{coarsening} $G$ by contracting edges or small clusters.
The resulting small graph $G'$ is then still a good representation of the overall input and an \emph{initial partition} of $G'$ already induces a good partition of $G$.
This is further improved by \emph{uncoarsening} the graph and improving the partition on each level through refinement algorithms.

However, parallelizing multi-level graph partitioning has proved challenging over several decades.
While shared-memory graph partitioners have recently matured to achieve high quality and reasonable scalability \cite{Mt-METIS, Mt-KaHIP, KaMinPar, Mt-KaHyPar-Q-F}, current distributed-memory partitioners \cite{ParHIP, ParMETIS, XtraPuLP} induce a severe quality deterioration and often are not able to consistently achieve feasible (i.e. balanced) partitions.
In particular, high quality partitioners do not scale to the number of processing elements (PEs) available in large supercomputers.
This situation is exacerbated by the fact that often the number of blocks $k$ should increase linearly in the number of PEs.
Previous systems are not able to directly handle large $k$ running into even larger problems with achieving feasibility.

In this paper, we present \algname\ which addresses all these issues.
Its basis is a distributed-memory adaptation of the deep-multilevel graph partitioning concept \cite{KaMinPar} that continues the multilevel approach deep into the initial partitioning phase. This makes the large $k$ case much easier and eliminates a parallelization bottleneck due to initial partitioning.
Our coarsening and refinement algorithms are based on the label propagation approach
previously used in several partitioners \cite{ParHIP,XtraPuLP,ParMETIS}.
Label propagation \cite{LPGeneral, DBLP:conf/wea/MeyerhenkeSS14} greedily moves vertices to other clusters/blocks when this reduces cuts (and does not violate the balance constraint). 
This is simple, fast, effective and robust even for complex networks. 
We develop a distributed-memory version with improved scalability, e.g., by using improved sparse-all-to-all primitives.
Perhaps the main algorithmic innovation are new scalable distributed techniques allowing to maintain the balance constraint.
During coarsening, a maximum cluster weight is approximated by unwinding contractions that
lead to overweight clusters. During uncoarsening, block weight constraints are achieved by finding, selecting and applying globally ``best'' block moves.

The experiments described in \Cref{s:experiments} indicate that our implementation has achieved
the main goals. It scales to at least \numprint{8192} cores even for complex networks
that did not scale on previous distributed solvers. Feasibility is guaranteed, even for large $k$ and
quality is typically within a few percent of the shared-memory systems.
\Cref{s:conclusion} summarizes the results and discusses possible future improvements.

\paragraph*{Contributions}
\begin{itemize}
\item Scalable distributed implementation of deep multilevel graph partitioning.
\item Simplicity using label propagation for both contraction and refinement.
\item New scalable balanced coarsening and uncoarsening algorithm.
\item Extensive evaluation on both large real world networks and huge synthetic networks from 3 input families.
\item Quality comparable to shared-memory systems.
\item Scalability up to (at least) $2^{13}$ cores and $2^{39}$ edges.
\item Works both for complex networks and large number of blocks
  where previous systems often fail.
\end{itemize}

\section{Preliminaries}\label{s:preliminaries}

\myparagraph{Notation and Definitions.}
Let $G = (V, E, c, \omega)$ be an undirected graph with vertex weights $c: V \rightarrow \mathbb{N}_{> 0}$, edge weights $\omega: E \rightarrow \mathbb{N}_{> 0}$, $n \coloneqq \lvert V \rvert$, and $m \coloneqq \lvert E \rvert$.
We extend $c$ and $\omega$ to sets, i.e., $c(V') \coloneqq \sum_{v \in V'} c(v)$ and $\omega(E') \coloneqq \sum_{e \in E'} \omega(e)$.
$N(v) \coloneqq \{ u \mid \{ u, v \} \in E\}$ denotes the neighbors of $v$.
For some $V' \subseteq V$, $G[V']$ denotes the subgraph of $G$ induced by $V'$.
We are looking for \emph{blocks} of nodes $\Pi \coloneqq \{ V_1, \dots, V_k \}$ that partition $V$, i.e., $V_1 \cup \dots \cup V_k = V$ and $V_i \cap V_j = \emptyset$ for $i \neq j$.
The \emph{balance constraint} demands that for all $i \in \{1, \dots, k\}$, $c(V_i) \le L_\text{max} \coloneqq \max \{ (1 + \varepsilon) \frac{c(V)}{k}, \frac{c(V)}{k} + \max_v c(v) \}$ for some imbalance parameter $\varepsilon$\footnote{Traditionally, $L_k \coloneqq (1 + \varepsilon) \lceil \frac{c(V)}{k} \rceil$ is used as balance constraint. We relax this constraint since it is otherwise NP-complete to find a feasible partition.}.
The objective is to minimize $\text{cut}(\Pi) \coloneqq \sum_{i < j} \omega(E_{ij})$ (weight of all cut edges), where $E_{ij} \coloneqq \{ \{u, v\} \in E \mid u \in V_i \text{ and } v \in V_j \}$.
We call a vertex $u \in V_i$ that has a neighbor in $V_j$, $i \neq j$, a \emph{boundary vertex}.
A \emph{clustering} $\mathcal{C} \coloneqq \{ C_1, \dots, C_\ell \}$ is also a partition of $V$, where the number of blocks $\ell$ is not given in advance (there is also no balance constraint).

\myparagraph{Machine Model and Input Format.}
The distributed memory model used in this work considers $P$ processing elements (PEs) numbered $1..P$, connected by a full-duplex, single ported communication network.
The input graph is given with a (usually balanced) 1D vertex partition.
Each PE is given a subgraph of the input graph (i.e., a block of the 1D partition) with consecutive vertices.
An undirected edge $\{u, v\}$ is represented by two directed edges $(u, v)$, $(v, u)$, which are stored on the PEs owning the respective tail vertices.
Vertices adjacent to vertices owned by other PEs are called \emph{interface vertices} and are replicated as \emph{ghost vertices} (i.e., without outgoing edges) on those PEs.

\section{Related Work}
There has been a huge amount of research on graph partitioning so that we refer the reader to overview papers \cite{GPOverviewBook, DBLP:reference/bdt/0003S19, RecentAdvances, MoreRecentAdvances} for most of the general material.
Here, we focus on parallel algorithms for high-quality graph partitioning.

\myparagraph{Distributed Graph Partitioning.}
Virtually all high-quality partitioners are based on the multilevel paradigm, e.g., \partitioner{ParMETIS}~\cite{ParMETIS, METIS},
\partitioner{ParHIP}~\cite{ParHIP, KaHIP}
and others~\cite{JOSTLE, PT-Scotch}.
These algorithms partition a graph in three phases.
First, they build a hierarchy of successively coarse approximations of the input graph, usually by contracting matchings or clusters. 
Once the graph has only few vertices left (e.g., $n \le C k$ for some \emph{contraction limit} $C$), the graph is partitioned into $k$ blocks.
Finally, this partition is successively projected onto finer levels of the hierarchy and refined using local search algorithms.

The performance of multilevel algorithms is defined by the algorithmic components used for these phases.
Partitioners designed for mesh-partitioning usually contract matchings to coarsen the graph \cite{ParMETIS, JOSTLE, PT-Scotch}.
However, this technique is not suitable for partitioning complex networks that only admit a small maximum matching.
Thus, other partitioners use two-hop matchings~\cite{METIS-2HopMatching} or size-constrained label propagation \cite{KaMinPar, Mt-KaHyPar-D, ParHIP}.
Due to its simple yet effective nature, the latter is also commonly used as a local search algorithm during refinement~\cite{ParHIP, ParMETIS, KaMinPar, Mt-KaHyPar-D, Mt-KaHIP, JOSTLE, DBLP:conf/ipps/DevineBHBC06}.

Label propagation has also been used by non-multilevel graph partitioning algorithms such as \partitioner{XtraPuLP}~\cite{XtraPuLP}, which reports scalability up to $2^{17}$ cores, a level which has not been reached by multilevel algorithms. 
However, using label propagation without the multilevel paradigm comes with a pronounced decline in quality; Ref. \cite{KaMinPar} reports edge cuts for \partitioner{PuLP}~\cite{PuLP} (non-multilevel) that are on average more than twice as large and those of \partitioner{KaMinPar} (multilevel).
Across a large and diverse benchmark set, this is considered a lot; most multilevel algorithms achieve average edge cuts within a few percentage points of each other.

Another class of highly scalable graph partitioners include geometric partitioners, which work on a geometric embedding of the graph.
While these algorithms are orders of magnitude faster than multilevel algorithms \cite{Geographer}, they generally compute larger edge cuts and only work on graphs with a meaningful geometric embedding.

\myparagraph{Deep Multilevel Graph Partitioning.}
As plain MGP algorithms usually shrink the graph down to $C k$ vertices, large values for $k$ break the assumption that the coarsest graph is small.
This causes their performance to deteriorate~\cite{KaMinPar}.
Instead, recursive bipartitioning can be used to compute partitions with large $k$, but this induces an additional $\log k$ factor in running time and makes it more difficult to compute balanced partitions due to the lack of global view.
\emph{Deep multilevel graph partitioning} (deep MGP)~\cite{KaMinPar} circumvents these problems by continuing coarsening deep into initial partitioning.
More precisely, deep MGP coarsens the graph until only $2C$ vertices are left, independent of $k$.
Hereby, parallelism is exploited by maintaining the invariant that each PE processes at least $C$ vertices, which is ensured by replicating coarser graphs and splitting the available PEs whenever the invariant would be violated.
After bipartitioning the coarsest graphs, it maintains the invariant that a (coarse) graph with $n$ vertices is partitioned into $\min\{k, n / C\}$ blocks by using recursive bipartitioning on the current level.
By using additional balancing techniques, partitioners based on deep MGP can obtain feasible high-quality partitions with a large number of blocks (e.g., $k \approx 1M$) while often being an order of magnitude faster than partitioners based on plain MGP.
Compared to recursive bipartitioning the entire graph, it reduces the additional $\log k$ factor to $\log kC/n$.
\partitioner{KaMinPar} \cite{KaMinPar} is a scalable shared-memory implementation of deep MGP which uses size-constrained label propagation during coarsening and refinement.

\section{Distributed Deep Multievel Graph Partitioning}\label{s:parallelization}

We introduce \algname, a distributed graph partitioner based on deep MGP.
We first describe the distributed deep MGP scheme itself, simplified by assuming that $k$ and $P$ are powers of two. 
Then, we outline the building blocks for coarsening, initial partitioning, refinement and balancing implemented in \algname.

\myparagraph{Distributed Deep Multilevel Graph Partitioning.}

\begin{figure*}[!t]
    \includegraphics[width=\textwidth]{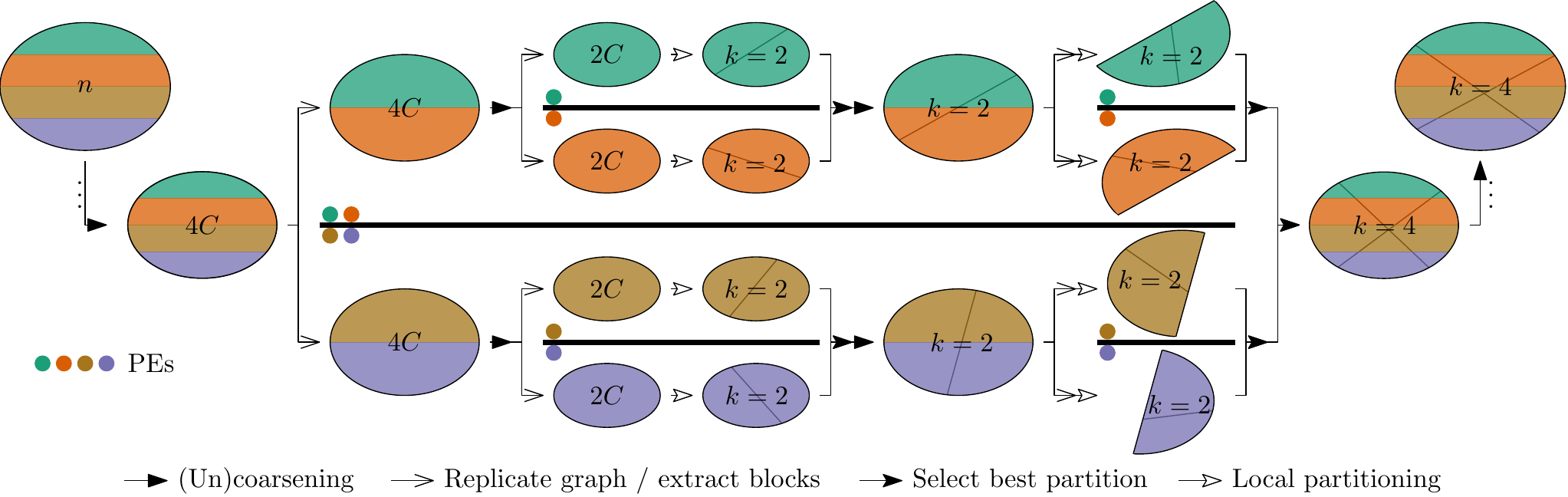}
        \caption{Distributed deep multilevel graph partitioning on $P = 4$ PEs. Unpartitioned graphs are labeled with their number of vertices, partitioned graphs are labeled with their number of blocks. Blocks are subdivided into $K = 2$ blocks, and the goal is to partition to graph into $k = 4$ blocks.}
    \label{fig:deep_mgp}
\end{figure*}

\ifarxiv
\begin{algorithm2e}[t]
    \KwIn{$G = (V, E)$, $k$, const. $C, K$}
    \KwOut{$k$-way partition $\Pi$ of $G$}
    \eIf(\tcp*[f]{deep coarsening}){$\lvert V(G) \rvert > C \cdot \min\{k, K\}$}{ \label{line:coarsening}
        \If(\tcp*[f]{duplicate graph if too small for $P$ PEs}){$\lvert V(G) \rvert < C \cdot P$}{ \label{line:replicationbegin}
            $c \coloneqq P / \mathrm{ceil}_2(\lvert V(G) \rvert / C)$ \tcp*[f]{number of copies} \\
            $G \coloneqq \FuncSty{Replicate}(G, c)$ \tcp*[f]{replicate graph $c$ times and group PEs} \\
            $P \coloneqq P / c$ \tcp*[f]{number of PEs per group} \\
        } \label{line:replicationend} 
        \tcp{standard multilevel graph partitioning:} 
        $G_c \coloneqq \FuncSty{Coarsen}(G)$ \label{line:coarsen} \\
        $\Pi_c \coloneqq \FuncSty{DeepMGP}(G_c, k, P)$ \label{line:partition} \\
        $\Pi \coloneqq \FuncSty{Project}(G, \Pi_c)$ \label{line:project} \\
        $\Pi \coloneqq \FuncSty{BalanceAndRefine}(G, \Pi)$ \label{line:refine} \\
    }(\tcp*[f]{base case}){ 
        $\Pi \coloneqq \{V\}$ \tcp*[f]{trivial $1$-way partition} \label{line:base} \\ 
    }
    $k' \coloneqq \min\{k, \textrm{ceil}_2(\lvert V \rvert / C)\}$ \\ 
    \While(\tcp*[f]{extend partition}){$\lvert \Pi \rvert < k'$}{ \label{line:extendbegin}
        $G'_1, \dots, G'_{\lvert \Pi \rvert / P} \coloneqq \FuncSty{DistributeBlocks}(G, \Pi)$ \tcp*[f]{$G'_i$s are local graphs} \label{line:distribute} \\
        \tcp{partition each local graph into $\min\{k' / \lvert \Pi \rvert, K\}$ blocks:}
        \For{$i \coloneqq 1$ \KwTo $\lvert \Pi \rvert / P$}{
                $\Pi'_i \coloneqq \FuncSty{LocalPartitioning}(G'_i, \min\{k' / \lvert \Pi \rvert, K\})$ \\
        }
        \tcp{combine local partitions $\Pi'_i$ of $G'_i$ to global partition $\Pi$ of $G$:}
        $\Pi \coloneqq \FuncSty{CollectPartitions}(\Pi'_1, \dots, \Pi'_{\lvert \Pi \rvert / P})$ \label{line:collect} \\
        $\Pi \coloneqq \FuncSty{BalanceAndRefine}(G, \Pi)$ \label{line:balance} \\
        } \label{line:extendend}
    \Return{$\Pi$}
    \caption{\normalsize $\FuncSty{DeepMGP}(G, k, P)$: Deep multilevel with $k$ blocks on $P$ PEs.}
    \label{alg:deep_mgp}
\end{algorithm2e}
\else
\fi

\ifarxiv
We outline the partitioning scheme in \Cref{fig:deep_mgp} and \Cref{alg:deep_mgp}.
\fi
Distributed deep MGP starts by coarsening the input graph down to $K \cdot C$ vertices, building a hierarchy of successively coarse graphs $G_1, \dots, G_{\ell}$
\ifarxiv
(\Cref{fig:deep_mgp}, left, and \Cref{alg:deep_mgp}, lines \ref{line:coarsen}--\ref{line:project}).
\fi
Here, $C$ is the \emph{contraction limit} and $K$ is a tuning parameter that generalizes the bipartitioning steps from Ref.~\cite{KaMinPar} to $K$-way partitioning.
To improve scalability on coarse levels, we follow Ref. \cite{JOSTLE} and maintain the invariant that $P$ PEs work on a graph with at least $P \cdot C$ vertices by splitting the PEs into groups and duplicating the current graph whenever this invariant would be violated 
\ifarxiv
otherwise (lines \ref{line:replicationbegin}--\ref{line:replicationend}).
\fi
Once coarsening converged, we gather the coarsest graph $G_{\ell}$ on all PEs and partition
it into $\min\{k, K\}$ blocks using a non-distributed graph partitioner (Figure~\ref{fig:deep_mgp}, middle).
The best partition (within each group of PEs) is selected and projected onto $G_{\ell - 1}$.
From here, we maintain two invariants: (1) the current partition is feasible, which is ensured by using the distributed balancing algorithm described below, and (2) a graph with $n_{\ell}$ vertices is partitioned into $\min\{k, \textrm{ceil}_2(n_{\ell} / C)\}$ blocks.
Here, $\textrm{ceil}_2(x)$ denotes the 
smallest power of $2$ equal to or larger than $x$.
The second invariant is maintained using recursive $K$-way partitioning.
More precisely, whenever the invariant is violated, we extract the block-induced subgraphs of the current partition and gather them on PEs such that each PE receives $k/P$ complete subgraphs (note that $k \ge P$ due to the duplication process described above).
From there, we use a non-distributed graph partitioner to recursively partition the subgraphs and project the new partitions back onto the distributed graph.
\ifarxiv
This process is illustrated by \Cref{alg:deep_mgp}, lines \ref{line:extendbegin}--\ref{line:extendend}.
\fi
The resulting partition, satisfying both invariants, is then improved using a distributed $k$-way refinement algorithm.
Note that if 
$k > \textrm{ceil}_2(\lvert V(G_1) \rvert / C)$, the partition computed on the finest graph has not enough blocks.
\ifarxiv
In this case, we distribute and partition the block-induced subgraphs once more to compute the missing blocks (omitted from \Cref{alg:deep_mgp}).
\fi

\myparagraph{Coarsening.}
We use a similar parallelization of size-constrained label propagation as \partitioner{ParHIP}~\cite{ParHIP} and \partitioner{KaMinPar}~\cite{KaMinPar}.
The algorithm works by first assigning each vertex to its own cluster.
In further iterations over the vertices (we use $\{3, 5\}$ iterations), they are then moved to adjacent clusters such that the weight of intra-cluster edges is maximized without violating the maximum cluster weight $W \coloneqq \varepsilon \frac{c(V)}{k'}$ where $k' \coloneqq \min\{k, \lvert V \rvert / C\}$~\cite{KaMinPar}.

As noted in Ref. \cite{DBLP:conf/wea/MeyerhenkeSS14, Mt-KaHIP}, the solution quality of label propagation is improved when iterating over vertices in increased degree order.
Since this is not cache efficient and lacks diversification by randomization, we sort the vertices into exponentially spaced degree buckets, i.e., bucket $i$ contains all vertices with degree $2^i \le d < 2^{i + 1}$, and rearrange the input graph accordingly. 
This happens locally on each PE, i.e., we do not sort the vertices globally.
Then, during label propagation, we split buckets into small chunks and randomize traversal on a inter-chunk and intra-chunk level.
This is analogous to the randomization of the matching algorithm used by \textsf{Metis}~\cite{METIS}.

To communicate the current cluster assignment of interface vertices, we follow \partitioner{ParHIP} and split each iteration into $\max\{\alpha, \beta / P\}$ (we use $\alpha = 8$, $\beta = 128$) batches.
After each batch, we use a sparse all-to-all operation to notify adjacent PEs of interface vertices that were moved to a different cluster.
Since clusters can span multiple PEs, enforcing the maximum cluster weight becomes more challenging than in a shared-memory setting.
\partitioner{ParHIP} relaxes the weight limit and only enforces it locally, i.e., allows clusters with weight up to $P \cdot W$.
This can lead to very heavy coarse vertices, making it more difficult to compute balanced partitions.
Instead, we track the global cluster weights by sending the change in cluster weight after each batch to the PE owning the initial vertex of the cluster, which accumulates the changes and replies with the total weight of the cluster.
If a cluster becomes heavier than $W$, each PE reverts moves proportional to its part of the total cluster weight. 
Those vertices can then be moved to other clusters during the next iteration.

After clustering the graph, we contract all clusters to build the next graph in the hierarchy.
We give more details on this operation in \Cref{s:implementation}.

\myparagraph{Refinement.}
We also use size-constrained label propagation to improve the current graph partition.
In contrast to label propagation for clustering as described above, vertices are initially assigned to clusters representing the blocks of the partition, and the maximum block weight is used as weight constraint.
We use the same iteration order and number of batches as during coarsening to move vertices to adjacent blocks such that the weight of intra-block edges is maximized without violating the balance constraint.
Ties are broken in favor of the lighter block, or by coin flip if both blocks have the same weight.

Since the number of blocks during refinement is usually much smaller than the number of clusters during coarsening, we track the global block weights using an allreduce operation after each vertex batch.
Note that this does not prevent violations of the balance constraint if multiple PEs move vertices to the same block during the same vertex batch.
In this case, we use our global balancing algorithm described below afterwards to restore the balance constraint.
This is a downside compared to refinement via size-constrained label propagation in shared-memory parallel graph partitioners, where the balance constraint can be strictly enforced using atomic compare-and-swap operations.

\myparagraph{Balancing.}
As noted in Ref. \cite{KaMinPar}, balance constraint violations during deep MGP can occur after initial partitioning or after projecting a coarse graph partition onto a finer level of the graph hierarchy.
Since these balance constraint violations are bounded by the weight of the heaviest vertex, we design the following balancing algorithm following the assumption that only few vertex moves are necessary to restore balance and that thus, it is feasible to invest a moderate amount of work per vertex movement.
The greedy algorithm works as follows.

For each overloaded block $B$, we maintain a priority queue $P_B$ of local vertices of that block on each PE.
The vertices in the priority queues are ordered by their \emph{relative gain}, which we define as $g \cdot c(v)$ if $g \ge 0$ and $g / c(v)$ if $g < 0$, where $g$ is the largest reduction in edge cut when moving $v$ to a block that would not become overloaded.
Note that this rating function generally prefers to move few heavy vertices over moving many lighter vertices, which follows our assumption that few vertex moves are sufficient to balance the partition.
To keep the priority queues small, we maintain the invariant that a priority queue $P_B$ stores no more vertices than are necessary to remove all excess weight $o(B) \coloneqq c(B) - L_\text{max}$ from $B$.
We initialize the priority queues by iterating over the vertices. 
If a vertex is in an overloaded block $B$ and $c(P_B) < o(B)$, we insert it. 
Otherwise, we only insert the vertex if its relative gain is larger than the lowest relative gain of any vertex in $P_B$ and remove its lowest vertex if $c(P_B) > o(B) + \max_v c(v)$ after insertion.

After initializing the priority queues, we use a binary tree reduction to repeatedly identify the $\ell$ highest scored vertices per block globally, for some constant input parameter $\ell$.
For each overloaded block, each PE contributes up to $\ell$ vertices from its local priority queue.
At each level of the reduction tree, the two lists of size $\le \ell$ are then merged and cut off after at most $\ell$ vertices, or sooner if a shorter prefix is sufficient to remove all excess weight from the corresponding block.
The root PE then decides which moves to perform such that no block becomes overloaded and broadcasts its decision to all PEs.
Using this information, PEs remove vertices that were moved from their priority queues and update the relative gain of neighbors of moved vertices.
We repeat this process until the partition is balanced.

\section{Implementation Details}\label{s:implementation}
\myparagraph{Vertex and Edge IDs.}
To reduce the communication overheads, we distinguish between local- and global vertex- and edge identifiers.
This allows us to use \SI{64}{bit} data types for global and \SI{32}{bit} data types for local IDs.

\myparagraph{Graph Contraction.}
Contracting a clustering consisting of $n_C$ clusters and constructing the corresponding coarse graph works as follows.
First, the clustering algorithm described above assigns a cluster ID to each vertex, which corresponds to some vertex ID in the distributed graph.
We say that a cluster is \emph{owned} by the PE owning the corresponding vertex.
After contracting the local subgraphs (i.e., deduplicating edges between clusters and accumulating vertex- and edge weights), we map clusters to PEs such that each PE gets roughly the same number of coarse vertices while attempting to minimize the required communication amount.
We assign $\le \delta \cdot n_C / P$ clusters owned by each PE to the same PE (in our experiments, $\delta = 1.1$).
If a PE owns more clusters, we redistribute the remaining clusters to PEs that have the smallest number of clusters assigned to them.
Afterwards, each PE sends outgoing edges of coarse vertices to the respective PE using an all-to-all operation, then builds the coarse graph by deduplicating edges and accumulating vertex- and edge weights.

\myparagraph{Low-latency Sparse All-to-All.}
Many steps of \algname\ require communication along the cut edges of the distributed graph, which translates to (often very) sparse and irregular all-to-all communication.
Since \texttt{MPI\_Alltoallv} has relatively high latency, we instead use a two-level approach that arranges PEs in a grid~\cite{MSTPaper}.
Then, messages are first sent to the right row, then to the right column, reducing the total number of messages send through the network from $\mathcal{O}(P^2)$ to $\mathcal{O}(P)$.

\section{Experiments}\label{s:experiments}

We implemented the proposed algorithm \algname\ in C++ and compiled it using g++-12.1 with flags \texttt{-O3 -march=native}.
We use OpenMPI 4.0 as parallelization library and \partitioner{growt}~\cite{growt} for hash tables.
The raw data of all experiments is available online\footnote{\url{https://algo2.iti.kit.edu/seemaier/ddeep_mgp/}}.

\myparagraph{Setup.}
We evaluate the solution quality of our algorithm on a shared-memory machine equipped with \SI{1}{TB} of main memory and one AMD EPYC 7702P processor with 64 cores (Machine~A). 
Additionally, we perform scalability experiments on a high-performance cluster where each compute node is equipped with \SI{256}{GB} of main memory and two Intel Xeon Platinum 8368 processors (Machine~B).
The compute nodes are connected by an InfiniBand 4X HDR \SI{200}{GBit/s} network with approx. \SI{1}{\micro s} latency.
We only use \numprint{64} out of the available \numprint{78} cores since some of the graph generators require the number of cores to be a power of two.
While our partitioner can use multiple threads per MPI process, we only evaluate the configuration with one thread per MPI process, since this usually gives the best performance.

We compare \algname\ against the distributed versions of the algorithms included in Ref.~\cite{KaMinPar}, i.e., \partitioner{ParHIP}~\cite{ParHIP} (v3.14) and \partitioner{ParMETIS}~\cite{ParMETIS} (v4.0.3).
We do not include the distributed version \partitioner{PuLP}~\cite{PuLP} (\partitioner{XtraPuLP}~\cite{XtraPuLP}) in our main comparison, since its quality is not competitive with multilevel partitioners.
\ifarxiv
Instead, a comparison against \partitioner{XtraPuLP} is available in \Cref{s:xtrapulp}.
\fi
We evaluate two configurations of our algorithm: \algnameFast\ uses $C = 2000$ as contraction limit (same as in Ref. \cite{KaMinPar}), \partitioner{KaMinPar}~\cite{KaMinPar} for initial partitioning and performs \numprint{3} iterations of label propagation during coarsening, whereas \algnameStrong\ uses $C = 5000$ (same as in Ref. \cite{ParHIP}), \partitioner{Mt-KaHyPar}~\cite{Mt-KaHyPar-D} for initial partitioning and \numprint{5} iterations of label propagation during coarsening. 

\myparagraph{Instances.}
We evaluate our algorithm on the graphs from benchmark set B of Ref. \cite{KaMinPar} and the graphs used in Ref. \cite{ParHIP}.
\ifarxiv
A list of all graphs is available in \Cref{tab:graphs}, \Cref{s:benchmark_instances}.
\fi
Additionally, we use the graph generator \partitioner{KaGen}~\cite{KaGen} to evaluate the scaling capabilities of our algorithm on huge randomly generated 2D and 3D geometric and hyperbolic graphs denoted \rggTwoD{N}{D}, \rggThreeD{N}{D} and \rhg{N}{D}{3.0}. 
These graphs have $2^N$ vertices per compute node (\numprint{64} cores) and average degree $D$. 
The random hyperbolic graphs have power-law exponent $3$.
The largest graphs of these families have $2^{33}$ vertices and $2^{39}$ edges. 

\myparagraph{Methodology.}
We call a combination of a graph and the number of blocks an \emph{instance}. 
For each instance, we perform 5 repetitions with different seeds and aggregate the edge cuts and running times using the arithmetic mean.
To aggregate over multiple instances, we use the geometric mean.

To compare the solution quality of different algorithms, we use \emph{performance profiles}~\cite{DBLP:journals/mp/DolanM02}.
Let $\mathcal{A}$ be the set of algorithms we want to compare, $\mathcal{I}$ the set of instances, and $q_A(I)$ the quality of algorithm $A \in \mathcal{A}$ on instance $I \in \mathcal{I}$.
For each algorithm $A$, we plot the fraction of instances $\frac{\lvert \mathcal{I}_A(\tau) \rvert}{\lvert \mathcal{I} \rvert}$ ($y$-axis) where $\mathcal{I}_A(\tau) \coloneqq \{ I \in \mathcal{I} \mid q_A(I) \le \tau \cdot \min_{A' \in \mathcal{A}} q_{A'}(I) \}$ and $\tau$ is on the $x$-axis.
Achieving higher fractions at lower $\tau$-values is considered better. 
For $\tau = 1$, the $y$-value indicates the percentage of instances for which an algorithm performs best. 

\myparagraph{Solution Quality and Running Time.}\label{s:expquality}

\begin{figure*}[t]
    \centering
    \input{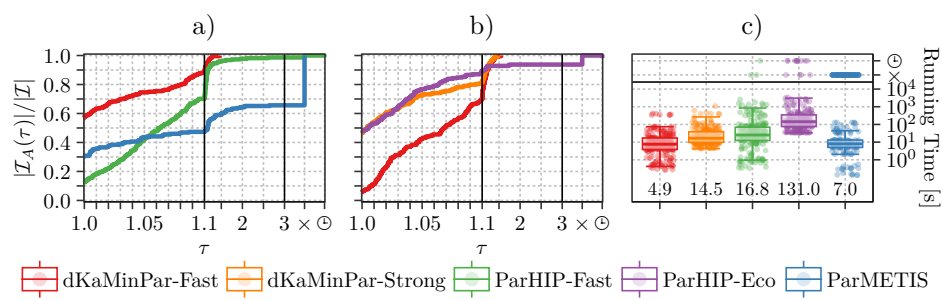}
\begin{tikzpicture}[x=1pt,y=1pt]
\definecolor{fillColor}{RGB}{255,255,255}
\begin{scope}
\definecolor{fillColor}{RGB}{255,255,255}

\path[fill=fillColor] ( 82.93,244.27) rectangle (100.27,261.62);
\end{scope}
\begin{scope}
\definecolor{drawColor}{RGB}{228,26,28}
\definecolor{fillColor}{RGB}{228,26,28}

\path[draw=drawColor,draw opacity=0.33,line width= 0.4pt,line join=round,line cap=round,fill=fillColor,fill opacity=0.33] ( 91.60,252.94) circle (  2.50);
\end{scope}
\begin{scope}
\definecolor{drawColor}{RGB}{228,26,28}

\path[draw=drawColor,line width= 0.6pt,line join=round] ( 84.66,252.94) -- ( 98.54,252.94);
\end{scope}
\begin{scope}
\definecolor{drawColor}{RGB}{228,26,28}

\path[draw=drawColor,line width= 0.6pt] ( 91.60,246.01) --
	( 91.60,248.61);

\path[draw=drawColor,line width= 0.6pt] ( 91.60,257.28) --
	( 91.60,259.88);
\definecolor{fillColor}{RGB}{255,255,255}

\path[draw=drawColor,line width= 0.6pt,fill=fillColor,fill opacity=0.50] ( 85.10,248.61) rectangle ( 98.11,257.28);

\path[draw=drawColor,line width= 0.6pt] ( 85.10,252.94) --
	( 98.11,252.94);
\end{scope}
\begin{scope}
\definecolor{fillColor}{RGB}{255,255,255}

\path[fill=fillColor] (159.58,244.27) rectangle (176.92,261.62);
\end{scope}
\begin{scope}
\definecolor{drawColor}{RGB}{255,127,0}
\definecolor{fillColor}{RGB}{255,127,0}

\path[draw=drawColor,draw opacity=0.33,line width= 0.4pt,line join=round,line cap=round,fill=fillColor,fill opacity=0.33] (168.25,252.94) circle (  2.50);
\end{scope}
\begin{scope}
\definecolor{drawColor}{RGB}{255,127,0}

\path[draw=drawColor,line width= 0.6pt,line join=round] (161.31,252.94) -- (175.19,252.94);
\end{scope}
\begin{scope}
\definecolor{drawColor}{RGB}{255,127,0}

\path[draw=drawColor,line width= 0.6pt] (168.25,246.01) --
	(168.25,248.61);

\path[draw=drawColor,line width= 0.6pt] (168.25,257.28) --
	(168.25,259.88);
\definecolor{fillColor}{RGB}{255,255,255}

\path[draw=drawColor,line width= 0.6pt,fill=fillColor,fill opacity=0.50] (161.74,248.61) rectangle (174.75,257.28);

\path[draw=drawColor,line width= 0.6pt] (161.74,252.94) --
	(174.75,252.94);
\end{scope}
\begin{scope}
\definecolor{fillColor}{RGB}{255,255,255}

\path[fill=fillColor] (244.66,244.27) rectangle (262.01,261.62);
\end{scope}
\begin{scope}
\definecolor{drawColor}{RGB}{77,175,74}
\definecolor{fillColor}{RGB}{77,175,74}

\path[draw=drawColor,draw opacity=0.33,line width= 0.4pt,line join=round,line cap=round,fill=fillColor,fill opacity=0.33] (253.33,252.94) circle (  2.50);
\end{scope}
\begin{scope}
\definecolor{drawColor}{RGB}{77,175,74}

\path[draw=drawColor,line width= 0.6pt,line join=round] (246.40,252.94) -- (260.27,252.94);
\end{scope}
\begin{scope}
\definecolor{drawColor}{RGB}{77,175,74}

\path[draw=drawColor,line width= 0.6pt] (253.33,246.01) --
	(253.33,248.61);

\path[draw=drawColor,line width= 0.6pt] (253.33,257.28) --
	(253.33,259.88);
\definecolor{fillColor}{RGB}{255,255,255}

\path[draw=drawColor,line width= 0.6pt,fill=fillColor,fill opacity=0.50] (246.83,248.61) rectangle (259.84,257.28);

\path[draw=drawColor,line width= 0.6pt] (246.83,252.94) --
	(259.84,252.94);
\end{scope}
\begin{scope}
\definecolor{fillColor}{RGB}{255,255,255}

\path[fill=fillColor] (307.47,244.27) rectangle (324.82,261.62);
\end{scope}
\begin{scope}
\definecolor{drawColor}{RGB}{152,78,163}
\definecolor{fillColor}{RGB}{152,78,163}

\path[draw=drawColor,draw opacity=0.33,line width= 0.4pt,line join=round,line cap=round,fill=fillColor,fill opacity=0.33] (316.15,252.94) circle (  2.50);
\end{scope}
\begin{scope}
\definecolor{drawColor}{RGB}{152,78,163}

\path[draw=drawColor,line width= 0.6pt,line join=round] (309.21,252.94) -- (323.08,252.94);
\end{scope}
\begin{scope}
\definecolor{drawColor}{RGB}{152,78,163}

\path[draw=drawColor,line width= 0.6pt] (316.15,246.01) --
	(316.15,248.61);

\path[draw=drawColor,line width= 0.6pt] (316.15,257.28) --
	(316.15,259.88);
\definecolor{fillColor}{RGB}{255,255,255}

\path[draw=drawColor,line width= 0.6pt,fill=fillColor,fill opacity=0.50] (309.64,248.61) rectangle (322.65,257.28);

\path[draw=drawColor,line width= 0.6pt] (309.64,252.94) --
	(322.65,252.94);
\end{scope}
\begin{scope}
\definecolor{fillColor}{RGB}{255,255,255}

\path[fill=fillColor] (368.90,244.27) rectangle (386.25,261.62);
\end{scope}
\begin{scope}
\definecolor{drawColor}{RGB}{55,126,184}
\definecolor{fillColor}{RGB}{55,126,184}

\path[draw=drawColor,draw opacity=0.33,line width= 0.4pt,line join=round,line cap=round,fill=fillColor,fill opacity=0.33] (377.57,252.94) circle (  2.50);
\end{scope}
\begin{scope}
\definecolor{drawColor}{RGB}{55,126,184}

\path[draw=drawColor,line width= 0.6pt,line join=round] (370.64,252.94) -- (384.51,252.94);
\end{scope}
\begin{scope}
\definecolor{drawColor}{RGB}{55,126,184}

\path[draw=drawColor,line width= 0.6pt] (377.57,246.01) --
	(377.57,248.61);

\path[draw=drawColor,line width= 0.6pt] (377.57,257.28) --
	(377.57,259.88);
\definecolor{fillColor}{RGB}{255,255,255}

\path[draw=drawColor,line width= 0.6pt,fill=fillColor,fill opacity=0.50] (371.07,248.61) rectangle (384.08,257.28);

\path[draw=drawColor,line width= 0.6pt] (371.07,252.94) --
	(384.08,252.94);
\end{scope}
\begin{scope}
\definecolor{drawColor}{RGB}{0,0,0}

\node[text=drawColor,anchor=base west,inner sep=0pt, outer sep=0pt, scale=  0.80] at (100.27,250.19) {dKaMinPar-Fast};
\end{scope}
\begin{scope}
\definecolor{drawColor}{RGB}{0,0,0}

\node[text=drawColor,anchor=base west,inner sep=0pt, outer sep=0pt, scale=  0.80] at (176.92,250.19) {dKaMinPar-Strong};
\end{scope}
\begin{scope}
\definecolor{drawColor}{RGB}{0,0,0}

\node[text=drawColor,anchor=base west,inner sep=0pt, outer sep=0pt, scale=  0.80] at (262.01,250.19) {ParHIP-Fast};
\end{scope}
\begin{scope}
\definecolor{drawColor}{RGB}{0,0,0}

\node[text=drawColor,anchor=base west,inner sep=0pt, outer sep=0pt, scale=  0.80] at (324.82,250.19) {ParHIP-Eco};
\end{scope}
\begin{scope}
\definecolor{drawColor}{RGB}{0,0,0}

\node[text=drawColor,anchor=base west,inner sep=0pt, outer sep=0pt, scale=  0.80] at (386.25,250.19) {ParMETIS};
\end{scope}
\end{tikzpicture}
    \caption{Results for $k = \{2, 4, 8, 16, 32, 64, 128\}$ with $\varepsilon = 3\%$ on Machine~A. From left to right: (a) edge cuts of \algnameFast, \partitioner{ParHIP-Fast} and \partitioner{ParMETIS}, (b) edge cuts of \algnameFast, \algnameStrong\ and \partitioner{ParHIP-Eco}, (c) running times of all algorithms. The numbers above the x-axis are geometric mean running times [s] over all instances for which all algorithms produced a result. Timeouts are marked with \symbTimeout, failed runs or infeasible results are marked with \symbInfeasible.}
    \label{fig:smallk}
\end{figure*}

We evaluate the quality and running time of \algname\ against competing distributed MGP algorithms using all \numprint{64} cores of Machine~A. 
Here, we partition the graphs of our benchmark set into $k \in \{2, 4, 8, 16, 32, 64, 128\}$ blocks with $\varepsilon = 3\%$.
To gain insights into the performance penalties of \algname\ due to its distributed nature, we also include a comparison against the shared-memory partitioner \partitioner{KaMinPar}.
\ifarxiv
Additionally, we report results for $k \in \{2^{11}, 2^{14}, 2^{17}, 2^{20}\}$ (\emph{large} $k$) in \Cref{s:exp:largek}.
\fi
We set the time limit for a single instance to one hour, which is approx. \numprint{10} times the running time of \algnameFast\ for small $k$ on most instances\footnote{Only \partitioner{twitter-2010} takes \SI{6}{\minute} resp. \SI{7}{\minute} for $k = 64$ resp. $k = 128$.}.

The results are summarized in \Cref{fig:smallk}a--c.
In \Cref{fig:smallk}a, we can see that \algnameFast\ finds the lowest edge cuts on approx. 60\% of all benchmark instances, whereas \partitioner{ParMETIS} and \partitioner{ParHIP-Fast} only find better edge cuts on approx. 30\% resp. 10\% of all instances.
Moreover, both competing algorithms frequently fail to compute feasible partitions --- in particular, \partitioner{ParMETIS} is unable to partition most social networks, violating the balance constraint or crashing on \R{\parmetisSmallkInfeasiblesPercentage\%} of all instances.
When looking at running times (\Cref{fig:smallk}c), we therefore only average over those instances for which all partitioners computed a feasible partition or ran into the timeout (\R{\smallkCommonRunningTimes} out of \R{\smallkInstances} instances).
\algnameFast\ (\R{\SI{\smallkCommonRunningTimedKaMinParFast}{\second}} geometric mean running time) is \R{\smallkCommonRunningTimeParMETISTodKaMinParFast} and \R{\smallkCommonRunningTimeParHIPFastTodKaMinParFast} times faster than \partitioner{ParMETIS} (\R{\SI{\smallkCommonRunningTimeParMETIS}{\second}}) and \partitioner{ParHIP-Fast} (\R{\SI{\smallkCommonRunningTimeParHIPFast}{\second}}), respectively. 
We attribute this to several reasons.
Compared to \partitioner{ParMETIS}, the lower running time is due to faster shrinkage of complex networks, while the advantage over \partitioner{ParHIP-Fast} is due to our more cache-efficient implementation of label propagation, more efficient graph contraction and the low-latency sparse all-to-all implementation described in~\Cref{s:implementation}.

ParHIP also offers a strong configuration \partitioner{ParHIP-Eco}, which performs more V-cycles and uses an evolutionary algorithm for initial partitioning to produce significantly better edge cuts at the cost of a much higher running time.
In \Cref{fig:smallk}b, we show that \algname\ can achieve the same solution quality when using the \algnameStrong\ configuration described above, while still being faster than \partitioner{ParHIP-Fast}.

\ifarxiv
\begin{figure*}[t]
    \centering
    \input{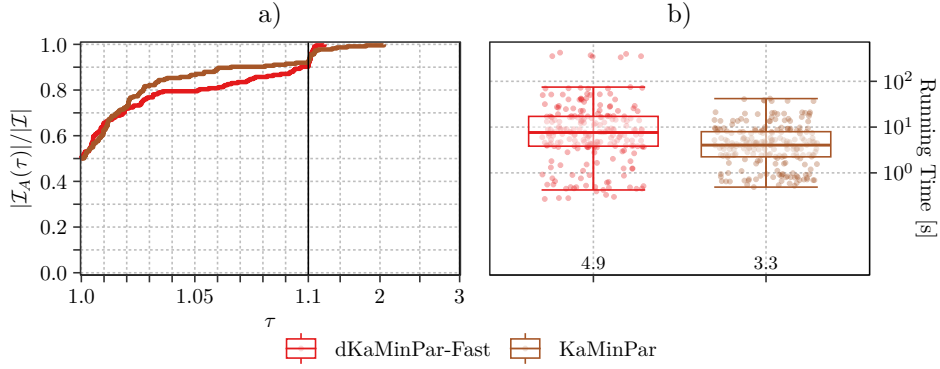}
\begin{tikzpicture}[x=1pt,y=1pt]
\definecolor{fillColor}{RGB}{255,255,255}
\begin{scope}
\definecolor{fillColor}{RGB}{255,255,255}

\path[fill=fillColor] (185.46,245.72) rectangle (199.91,260.17);
\end{scope}
\begin{scope}
\definecolor{drawColor}{RGB}{228,26,28}
\definecolor{fillColor}{RGB}{228,26,28}

\path[draw=drawColor,draw opacity=0.33,line width= 0.4pt,line join=round,line cap=round,fill=fillColor,fill opacity=0.33] (192.68,252.94) circle (  0.89);
\end{scope}
\begin{scope}
\definecolor{drawColor}{RGB}{228,26,28}

\path[draw=drawColor,line width= 0.6pt,line join=round] (186.90,252.94) -- (198.47,252.94);
\end{scope}
\begin{scope}
\definecolor{drawColor}{RGB}{228,26,28}

\path[draw=drawColor,line width= 0.6pt] (192.68,247.16) --
	(192.68,249.33);

\path[draw=drawColor,line width= 0.6pt] (192.68,256.56) --
	(192.68,258.73);
\definecolor{fillColor}{RGB}{255,255,255}

\path[draw=drawColor,line width= 0.6pt,fill=fillColor,fill opacity=0.50] (187.26,249.33) rectangle (198.11,256.56);

\path[draw=drawColor,line width= 0.6pt] (187.26,252.94) --
	(198.11,252.94);
\end{scope}
\begin{scope}
\definecolor{fillColor}{RGB}{255,255,255}

\path[fill=fillColor] (269.41,245.72) rectangle (283.86,260.17);
\end{scope}
\begin{scope}
\definecolor{drawColor}{RGB}{166,86,40}
\definecolor{fillColor}{RGB}{166,86,40}

\path[draw=drawColor,draw opacity=0.33,line width= 0.4pt,line join=round,line cap=round,fill=fillColor,fill opacity=0.33] (276.64,252.94) circle (  0.89);
\end{scope}
\begin{scope}
\definecolor{drawColor}{RGB}{166,86,40}

\path[draw=drawColor,line width= 0.6pt,line join=round] (270.85,252.94) -- (282.42,252.94);
\end{scope}
\begin{scope}
\definecolor{drawColor}{RGB}{166,86,40}

\path[draw=drawColor,line width= 0.6pt] (276.64,247.16) --
	(276.64,249.33);

\path[draw=drawColor,line width= 0.6pt] (276.64,256.56) --
	(276.64,258.73);
\definecolor{fillColor}{RGB}{255,255,255}

\path[draw=drawColor,line width= 0.6pt,fill=fillColor,fill opacity=0.50] (271.22,249.33) rectangle (282.06,256.56);

\path[draw=drawColor,line width= 0.6pt] (271.22,252.94) --
	(282.06,252.94);
\end{scope}
\begin{scope}
\definecolor{drawColor}{RGB}{0,0,0}

\node[text=drawColor,anchor=base west,inner sep=0pt, outer sep=0pt, scale=  0.80] at (205.41,250.19) {dKaMinPar-Fast};
\end{scope}
\begin{scope}
\definecolor{drawColor}{RGB}{0,0,0}

\node[text=drawColor,anchor=base west,inner sep=0pt, outer sep=0pt, scale=  0.80] at (289.36,250.19) {KaMinPar};
\end{scope}
\end{tikzpicture}
    \caption{Results for $k = \{2, 4, 8, 16, 32, 64, 128\}$ with $\varepsilon = 3\%$ on Machine~A. From left to right: (a) edge
cuts of \algnameFast\ and \partitioner{KaMinPar}, (b) corresponding running times. The numbers above the x-axis are geometric mean running times [s].}
    \label{fig:exp:kaminpar}
\end{figure*}
\fi

Finally, we compare out distributed partitioner against \partitioner{KaMinPar}, which is a shared-memory implementation of deep multilevel graph partitioning with label propagation for coarsening and refinement. 
\ifarxiv
The results are summarized in \Cref{fig:exp:kaminpar}.
\fi
The edge cut quality of both partitioners is almost the same, with a difference in average edge cut computed of less than 0.5\%.
However, the distributed partitioner is approx. 50\% slower than the shared-memory partitioner. 
This is expected, as communication through message passing is generally less efficient than shared-memory communication.

\myparagraph{Weak Scalability of \algname.}\label{s:expscalability}

\begin{figure*}[t]
    \centering 
    \input{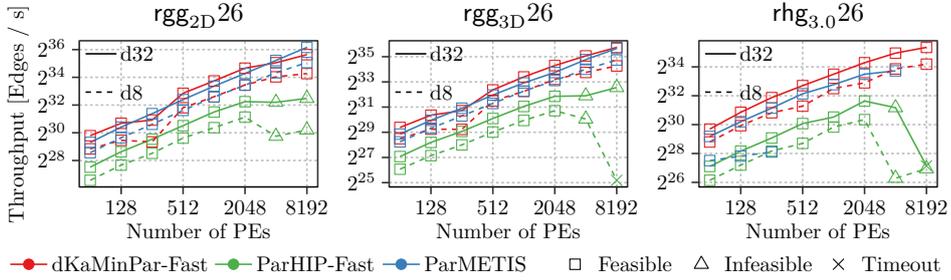}
\begin{tikzpicture}[x=1pt,y=1pt]
\definecolor{fillColor}{RGB}{255,255,255}
\begin{scope}
\definecolor{fillColor}{RGB}{255,255,255}

\path[fill=fillColor] ( 78.21,244.27) rectangle ( 95.56,261.62);
\end{scope}
\begin{scope}
\definecolor{drawColor}{RGB}{228,26,28}

\path[draw=drawColor,line width= 0.6pt,line join=round] ( 79.95,252.94) -- ( 93.82,252.94);
\end{scope}
\begin{scope}
\definecolor{drawColor}{RGB}{228,26,28}
\definecolor{fillColor}{RGB}{228,26,28}

\path[draw=drawColor,line width= 0.4pt,line join=round,line cap=round,fill=fillColor] ( 86.89,252.94) circle (  1.96);
\end{scope}
\begin{scope}
\definecolor{fillColor}{RGB}{255,255,255}

\path[fill=fillColor] (154.86,244.27) rectangle (172.20,261.62);
\end{scope}
\begin{scope}
\definecolor{drawColor}{RGB}{77,175,74}

\path[draw=drawColor,line width= 0.6pt,line join=round] (156.59,252.94) -- (170.47,252.94);
\end{scope}
\begin{scope}
\definecolor{drawColor}{RGB}{77,175,74}
\definecolor{fillColor}{RGB}{77,175,74}

\path[draw=drawColor,line width= 0.4pt,line join=round,line cap=round,fill=fillColor] (163.53,252.94) circle (  1.96);
\end{scope}
\begin{scope}
\definecolor{fillColor}{RGB}{255,255,255}

\path[fill=fillColor] (217.67,244.27) rectangle (235.01,261.62);
\end{scope}
\begin{scope}
\definecolor{drawColor}{RGB}{55,126,184}

\path[draw=drawColor,line width= 0.6pt,line join=round] (219.40,252.94) -- (233.28,252.94);
\end{scope}
\begin{scope}
\definecolor{drawColor}{RGB}{55,126,184}
\definecolor{fillColor}{RGB}{55,126,184}

\path[draw=drawColor,line width= 0.4pt,line join=round,line cap=round,fill=fillColor] (226.34,252.94) circle (  1.96);
\end{scope}
\begin{scope}
\definecolor{drawColor}{RGB}{0,0,0}

\node[text=drawColor,anchor=base west,inner sep=0pt, outer sep=0pt, scale=  0.80] at ( 95.56,250.19) {dKaMinPar-Fast};
\end{scope}
\begin{scope}
\definecolor{drawColor}{RGB}{0,0,0}

\node[text=drawColor,anchor=base west,inner sep=0pt, outer sep=0pt, scale=  0.80] at (172.20,250.19) {ParHIP-Fast};
\end{scope}
\begin{scope}
\definecolor{drawColor}{RGB}{0,0,0}

\node[text=drawColor,anchor=base west,inner sep=0pt, outer sep=0pt, scale=  0.80] at (235.01,250.19) {ParMETIS};
\end{scope}
\begin{scope}
\definecolor{fillColor}{RGB}{255,255,255}

\path[fill=fillColor] (282.73,244.27) rectangle (300.07,261.62);
\end{scope}
\begin{scope}
\definecolor{drawColor}{RGB}{0,0,0}

\path[draw=drawColor,line width= 0.4pt,line join=round,line cap=round] (289.44,250.98) rectangle (293.36,254.91);
\end{scope}
\begin{scope}
\definecolor{fillColor}{RGB}{255,255,255}

\path[fill=fillColor] (330.02,244.27) rectangle (347.36,261.62);
\end{scope}
\begin{scope}
\definecolor{drawColor}{RGB}{0,0,0}

\path[draw=drawColor,line width= 0.4pt,line join=round,line cap=round] (338.69,256.00) --
	(341.33,251.42) --
	(336.05,251.42) --
	cycle;
\end{scope}
\begin{scope}
\definecolor{fillColor}{RGB}{255,255,255}

\path[fill=fillColor] (381.33,244.27) rectangle (398.67,261.62);
\end{scope}
\begin{scope}
\definecolor{drawColor}{RGB}{0,0,0}

\path[draw=drawColor,line width= 0.4pt,line join=round,line cap=round] (388.04,250.98) -- (391.96,254.91);

\path[draw=drawColor,line width= 0.4pt,line join=round,line cap=round] (388.04,254.91) -- (391.96,250.98);
\end{scope}
\begin{scope}
\definecolor{drawColor}{RGB}{0,0,0}

\node[text=drawColor,anchor=base west,inner sep=0pt, outer sep=0pt, scale=  0.80] at (300.07,250.19) {Feasible};
\end{scope}
\begin{scope}
\definecolor{drawColor}{RGB}{0,0,0}

\node[text=drawColor,anchor=base west,inner sep=0pt, outer sep=0pt, scale=  0.80] at (347.36,250.19) {Infeasible};
\end{scope}
\begin{scope}
\definecolor{drawColor}{RGB}{0,0,0}

\node[text=drawColor,anchor=base west,inner sep=0pt, outer sep=0pt, scale=  0.80] at (398.67,250.19) {Timeout};
\end{scope}
\end{tikzpicture}
    \caption{Throughput of $\textsf{rgg}_{\textrm{2D}}$, $\textsf{rgg}_{\textrm{3D}}$ and \textsf{rhg} graphs with $2^{26}$ vertices per compute node, average degree $\in \{8, 32\}$, $k = 16$ and $\varepsilon = 3\%$ on \numprint{64}--\numprint{8192} cores of Machine~B.}
    \label{fig:ws_smallk}
\end{figure*}

We evaluate the weak scalability of \algname\ using \numprint{64}--\numprint{8192} cores (i.e., \numprint{1}--\numprint{128} compute nodes) of the high-performance cluster Machine~B.
For benchmark instances, we use randomly generated geometric (2D and 3D) and hyperbolic graphs with $2^{26}$ vertices per compute node and average degree $\in \{8, 32\}$.

In \Cref{fig:ws_smallk}, we partition these graphs into $k = 16$ blocks and observe weak scalability for \algnameFast\ all the way to \numprint{8192} cores on all three graph families.
On the geometric graphs, we achieve similar throughputs to \partitioner{ParMETIS}, while \partitioner{ParHIP-Fast} shows a drop in scalability beyond \numprint{2048} cores.
This is most likely due to the extensive and inefficient communication performed by \partitioner{ParHIP-Fast} during graph contraction.
Moreover, we note that \partitioner{ParHIP-Fast} was originally designed to overlap local work and global communication during label propagation through the use of nonblocking MPI operations.
This implementation relies on MPI progression threads, which seem to be unavailable in modern OpenMPI versions.
\partitioner{ParMETIS} shows significantly worse throughputs on random hyperbolic graphs and is unable to compute a partition on \numprint{8192} cores.
As mentioned before, this is most likely due to its inefficient coarsening strategy on graphs that follow a power-law degree distribution.

\ifarxiv
Looking at edge cuts (\Cref{tab:ws_cuts} in \Cref{s:ws_cuts}, upper half), \partitioner{ParMETIS} finds lower edge cuts than \algnameFast\ on the dense $\rggTwoD{26}{32}$ graph and both $\textsf{rgg}_{\textrm{3D}}$ graphs by 5\%--13\%.
\fi
However, on the sparser $\rggTwoD{26}{8}$ graph, \algnameFast\ has 19\% smaller cuts than \partitioner{ParMETIS} which is already a considerable improvement. 
The gap gets much larger for the hyperbolic graph where \partitioner{ParMETIS} only finds approx. 5.5--6.1 times larger cuts. 
Such solutions will be unsuitable for many applications.

\begin{figure*}[t]
    \centering 
    \input{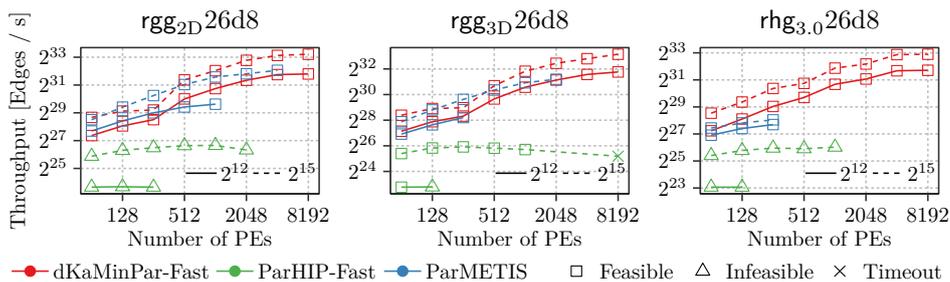}
\begin{tikzpicture}[x=1pt,y=1pt]
\definecolor{fillColor}{RGB}{255,255,255}
\begin{scope}
\definecolor{fillColor}{RGB}{255,255,255}

\path[fill=fillColor] ( 78.21,244.27) rectangle ( 95.56,261.62);
\end{scope}
\begin{scope}
\definecolor{drawColor}{RGB}{228,26,28}

\path[draw=drawColor,line width= 0.6pt,line join=round] ( 79.95,252.94) -- ( 93.82,252.94);
\end{scope}
\begin{scope}
\definecolor{drawColor}{RGB}{228,26,28}
\definecolor{fillColor}{RGB}{228,26,28}

\path[draw=drawColor,line width= 0.4pt,line join=round,line cap=round,fill=fillColor] ( 86.89,252.94) circle (  1.96);
\end{scope}
\begin{scope}
\definecolor{fillColor}{RGB}{255,255,255}

\path[fill=fillColor] (154.86,244.27) rectangle (172.20,261.62);
\end{scope}
\begin{scope}
\definecolor{drawColor}{RGB}{77,175,74}

\path[draw=drawColor,line width= 0.6pt,line join=round] (156.59,252.94) -- (170.47,252.94);
\end{scope}
\begin{scope}
\definecolor{drawColor}{RGB}{77,175,74}
\definecolor{fillColor}{RGB}{77,175,74}

\path[draw=drawColor,line width= 0.4pt,line join=round,line cap=round,fill=fillColor] (163.53,252.94) circle (  1.96);
\end{scope}
\begin{scope}
\definecolor{fillColor}{RGB}{255,255,255}

\path[fill=fillColor] (217.67,244.27) rectangle (235.01,261.62);
\end{scope}
\begin{scope}
\definecolor{drawColor}{RGB}{55,126,184}

\path[draw=drawColor,line width= 0.6pt,line join=round] (219.40,252.94) -- (233.28,252.94);
\end{scope}
\begin{scope}
\definecolor{drawColor}{RGB}{55,126,184}
\definecolor{fillColor}{RGB}{55,126,184}

\path[draw=drawColor,line width= 0.4pt,line join=round,line cap=round,fill=fillColor] (226.34,252.94) circle (  1.96);
\end{scope}
\begin{scope}
\definecolor{drawColor}{RGB}{0,0,0}

\node[text=drawColor,anchor=base west,inner sep=0pt, outer sep=0pt, scale=  0.80] at ( 95.56,250.19) {dKaMinPar-Fast};
\end{scope}
\begin{scope}
\definecolor{drawColor}{RGB}{0,0,0}

\node[text=drawColor,anchor=base west,inner sep=0pt, outer sep=0pt, scale=  0.80] at (172.20,250.19) {ParHIP-Fast};
\end{scope}
\begin{scope}
\definecolor{drawColor}{RGB}{0,0,0}

\node[text=drawColor,anchor=base west,inner sep=0pt, outer sep=0pt, scale=  0.80] at (235.01,250.19) {ParMETIS};
\end{scope}
\begin{scope}
\definecolor{fillColor}{RGB}{255,255,255}

\path[fill=fillColor] (282.73,244.27) rectangle (300.07,261.62);
\end{scope}
\begin{scope}
\definecolor{drawColor}{RGB}{0,0,0}

\path[draw=drawColor,line width= 0.4pt,line join=round,line cap=round] (289.44,250.98) rectangle (293.36,254.91);
\end{scope}
\begin{scope}
\definecolor{fillColor}{RGB}{255,255,255}

\path[fill=fillColor] (330.02,244.27) rectangle (347.36,261.62);
\end{scope}
\begin{scope}
\definecolor{drawColor}{RGB}{0,0,0}

\path[draw=drawColor,line width= 0.4pt,line join=round,line cap=round] (338.69,256.00) --
	(341.33,251.42) --
	(336.05,251.42) --
	cycle;
\end{scope}
\begin{scope}
\definecolor{fillColor}{RGB}{255,255,255}

\path[fill=fillColor] (381.33,244.27) rectangle (398.67,261.62);
\end{scope}
\begin{scope}
\definecolor{drawColor}{RGB}{0,0,0}

\path[draw=drawColor,line width= 0.4pt,line join=round,line cap=round] (388.04,250.98) -- (391.96,254.91);

\path[draw=drawColor,line width= 0.4pt,line join=round,line cap=round] (388.04,254.91) -- (391.96,250.98);
\end{scope}
\begin{scope}
\definecolor{drawColor}{RGB}{0,0,0}

\node[text=drawColor,anchor=base west,inner sep=0pt, outer sep=0pt, scale=  0.80] at (300.07,250.19) {Feasible};
\end{scope}
\begin{scope}
\definecolor{drawColor}{RGB}{0,0,0}

\node[text=drawColor,anchor=base west,inner sep=0pt, outer sep=0pt, scale=  0.80] at (347.36,250.19) {Infeasible};
\end{scope}
\begin{scope}
\definecolor{drawColor}{RGB}{0,0,0}

\node[text=drawColor,anchor=base west,inner sep=0pt, outer sep=0pt, scale=  0.80] at (398.67,250.19) {Timeout};
\end{scope}
\end{tikzpicture}
    \caption{Throughput of \textsf{rgg2D}, \textsf{rgg3D} and \textsf{rhg} graphs with $2^{26}$ vertices per compute node, average degree $8$, and $\varepsilon = 3\%$ on \numprint{64}--\numprint{8192} cores of Machine~B.
        The number of blocks are scaled with the size of the graph such that each block contains $2^{12}$ or $2^{15}$ vertices.}
    \label{fig:ws_largek}
\end{figure*}

We now evaluate weak scalability in terms of graph size \emph{and} number of blocks by scaling $k$ with the number of compute nodes used. 
This implies that the number of blocks is large when using a large number of cores.
The throughput of each algorithm in this setting is summarized in \Cref{fig:ws_largek}.
Note that we only use the sparser graphs in this experiment, since \partitioner{ParMETIS} and \partitioner{ParHIP} are unable to partition the dense versions of the graphs even on few compute nodes.

\partitioner{ParHIP-Fast} is unable to obtain a feasible partition on all but \numprint{6} instances, none of which uses more than \numprint{1024} cores, and only shows increasing throughputs up to \numprint{256} cores.
While \partitioner{ParMETIS} achieves decent weak scalability and computes feasible solutions on the mesh-type graphs, it is unable to partition any graph on \numprint{8192} cores and often crashes on fewer cores (e.g., it only works on up to \numprint{1024} cores on $\textsf{rgg}_{\textrm{2D}}$ with $2^{12}$ vertices per block).
On the random hyperbolic graph, it only computes a feasible solution on \numprint{64} cores.
Meanwhile, \algnameFast\ shows weak scalability up to \numprint{8192} cores \emph{on every graph family}, although it should be noted that the throughput increase from \numprint{4096} to \numprint{8192} is rather small.

In terms of number of edges cut, we summarize that \algname\ finds on average \R{\wsLargekdKaMinParToParMETISPercent\%} and \R{\wsLargekdKaMinParToParHIPPercent\%} lower edge cuts than \partitioner{ParMETIS} and \partitioner{ParHIP-Fast}, respectively (only averaging over instances for which the respective partitioner computed a feasible partition), with improvements ranging from 0\% on \rggThreeD{26}{8} to approx. 60\% on \rhg{3.0}{26}{8} ($2^{15}$ vertices per block).
\ifarxiv
Detailed per-instance edge cut results are available in \Cref{tab:ws_cuts} (\Cref{s:ws_cuts}).
\fi

\begin{figure}[t]
    \centering 
    \input{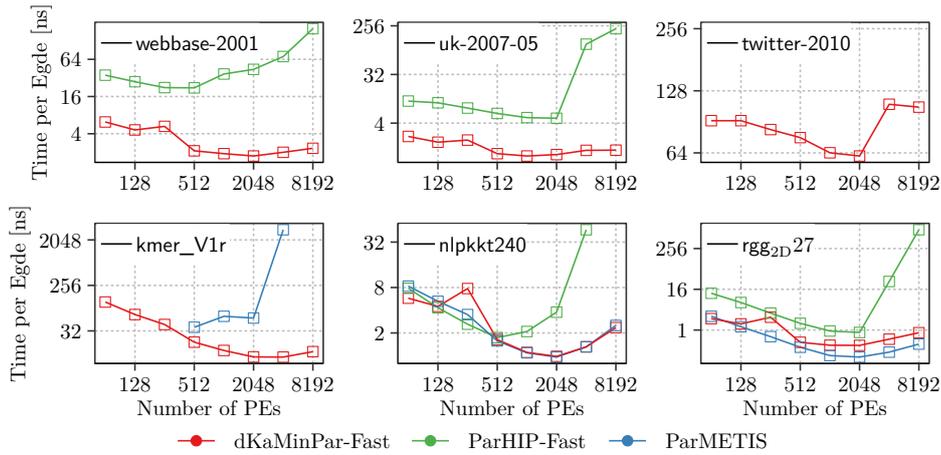}
\begin{tikzpicture}[x=1pt,y=1pt]
\definecolor{fillColor}{RGB}{255,255,255}
\begin{scope}
\definecolor{fillColor}{RGB}{255,255,255}

\path[fill=fillColor] (145.19,244.27) rectangle (162.53,261.62);
\end{scope}
\begin{scope}
\definecolor{drawColor}{RGB}{228,26,28}

\path[draw=drawColor,line width= 0.6pt,line join=round] (146.92,252.94) -- (160.80,252.94);
\end{scope}
\begin{scope}
\definecolor{drawColor}{RGB}{228,26,28}
\definecolor{fillColor}{RGB}{228,26,28}

\path[draw=drawColor,line width= 0.4pt,line join=round,line cap=round,fill=fillColor] (153.86,252.94) circle (  1.96);
\end{scope}
\begin{scope}
\definecolor{fillColor}{RGB}{255,255,255}

\path[fill=fillColor] (232.83,244.27) rectangle (250.18,261.62);
\end{scope}
\begin{scope}
\definecolor{drawColor}{RGB}{77,175,74}

\path[draw=drawColor,line width= 0.6pt,line join=round] (234.57,252.94) -- (248.44,252.94);
\end{scope}
\begin{scope}
\definecolor{drawColor}{RGB}{77,175,74}
\definecolor{fillColor}{RGB}{77,175,74}

\path[draw=drawColor,line width= 0.4pt,line join=round,line cap=round,fill=fillColor] (241.51,252.94) circle (  1.96);
\end{scope}
\begin{scope}
\definecolor{fillColor}{RGB}{255,255,255}

\path[fill=fillColor] (306.64,244.27) rectangle (323.99,261.62);
\end{scope}
\begin{scope}
\definecolor{drawColor}{RGB}{55,126,184}

\path[draw=drawColor,line width= 0.6pt,line join=round] (308.38,252.94) -- (322.25,252.94);
\end{scope}
\begin{scope}
\definecolor{drawColor}{RGB}{55,126,184}
\definecolor{fillColor}{RGB}{55,126,184}

\path[draw=drawColor,line width= 0.4pt,line join=round,line cap=round,fill=fillColor] (315.32,252.94) circle (  1.96);
\end{scope}
\begin{scope}
\definecolor{drawColor}{RGB}{0,0,0}

\node[text=drawColor,anchor=base west,inner sep=0pt, outer sep=0pt, scale=  0.80] at (168.03,250.19) {dKaMinPar-Fast};
\end{scope}
\begin{scope}
\definecolor{drawColor}{RGB}{0,0,0}

\node[text=drawColor,anchor=base west,inner sep=0pt, outer sep=0pt, scale=  0.80] at (255.68,250.19) {ParHIP-Fast};
\end{scope}
\begin{scope}
\definecolor{drawColor}{RGB}{0,0,0}

\node[text=drawColor,anchor=base west,inner sep=0pt, outer sep=0pt, scale=  0.80] at (329.49,250.19) {ParMETIS};
\end{scope}
\end{tikzpicture}
    \caption{Strong scaling running times for the largest low- and high-degree graphs in our benchmark set, with $k = 16$, $\varepsilon = 3\%$ on \numprint{64}--\numprint{8192} cores of Machine~B.}
    \label{fig:ss}
\end{figure}

\myparagraph{Strong Scalability of \algname.}
We now look at strong scalability of \algname. 
Here, we partition three of the largest low- and high-degree graphs from our benchmark set into $k = 16$ blocks using \numprint{64}--\numprint{8192} cores of Machine~B and a time limit of \SI{15}{\minute}.
The results are summarized in \Cref{fig:ss}, where we can observe strong scalability for up to \numprint{1024}--\numprint{2048} cores on high-degree graphs. 
\partitioner{ParMETIS} is unable to partition these graphs regardless of the number of cores used.
While \partitioner{ParHIP-Fast} scales up to \numprint{2048} cores on \partitioner{uk-2007-05}, it should be noted that its running time is still higher than \algname\ on just \numprint{64} cores.
The twitter graph is difficult to coarsen due to its highly skewed degree distribution; here, we observe that only \algname\ can partition the graph within the time limit.

Turning towards graphs with small maximum degree, we observe strong scalability for up to \numprint{2048}, \numprint{2048} and \numprint{1024} cores on \textsf{kmer\_V1r}, \textsf{nlpkkt240} and $\textsf{rgg}_{\textrm{2D}}27$, respectively.
Other algorithms seem to be unable to partition \textsf{kmer\_V1r}, which \partitioner{ParMETIS} is only able to partition on \numprint{512}--\numprint{4096} cores, even though the graph is relatively small and fits into the memory of a single compute node.
\partitioner{ParHIP-Fast} is not able to partition the graph at all.
Similar to our weak scaling experiments, \partitioner{ParMETIS} shows better scalability and throughput on the mesh-type graph $\textsf{rgg}_{\textrm{2D}}$ as well as on \partitioner{nlpkkt240}.

\ifarxiv
The edge cuts obtained remain relatively constant (\Cref{tab:ss_cuts} in \Cref{s:ss_cuts}) when scaling to large number of cores.
\fi
Surprisingly, the geometric mean edge cut on \numprint{8192} cores is slightly better than on \numprint{64} cores (by \R{\ssdKaMinParGmeanDecline\%}).

\section{Conclusion and Future Work}\label{s:conclusion}
Our distributed-memory graph partitioner \algname\ successfully partitions a wide range of input graphs using many thousands of cores yielding high speed and good quality.
Further improvements of the implementation might be possible, for example making better use of shared-memory on each compute node. 
Beyond that, one can explore the quality versus time trade off. 
By distributed implementations of more powerful local improvement algorithms like local search or flow-based techniques one could achieve better quality at the price of higher execution time.
It then also makes sense to look at a portfolio of different partitioners variants that can be run in parallel achieving good quality for subsets of inputs. 
For example, matching based coarsening as in \partitioner{ParMETIS} might help for mesh-like networks. 
On the other hand, more aggressive methods for handling high-degree nodes might help with some social networks.

\bibliographystyle{splncs04}
\bibliography{references}

\ifarxiv
\newpage
\section{Benchmark Instances}\label{s:benchmark_instances}
\begin{table}[h]
    \centering
    \caption{Basic properties of the benchmark set. Graphs are roughly classified as \emph{low degree} and \emph{high degree} graphs based on their maximum degree $\Delta$.}
    \begin{tabular}{c|l|rrr|r}
            \multicolumn{1}{c}{} & Graph & $n$ & $m$ & $\Delta$ & Ref. \\
        \midrule
            \multirow{17}{*}{\rotatebox[origin = c]{90}{Low-degree graphs}}
            & \textsf{packing} & \numprint{2145839} & \numprint{34976486} & \numprint{18} & \cite{DIMACS2012} \\
            & \textsf{channel} & \numprint{4802000} & \numprint{85362744} & \numprint{18} & \cite{DIMACS2012} \\
            & \textsf{hugebubbles} & \numprint{19458087} & \numprint{58359528} & \numprint{3} & \cite{DIMACS2012} \\
            & \textsf{nlpkkt240} & \numprint{27993600} & \numprint{746478752} & \numprint{27} & \cite{DIMACS2012} \\
            & \textsf{europe.osm} & \numprint{50912018} & \numprint{108109320} & \numprint{13} & \cite{DIMACS2012} \\
            & \textsf{kmerU1a} & \numprint{64678340} & \numprint{132787258} & \numprint{35} & \cite{FSMC} \\
            & $\textsf{rgg}_{\textrm{3D}}26$ & \numprint{67106449} & \numprint{755904090} & \numprint{34} & \cite{KaGen} \\
            & $\textsf{rgg}_{\textrm{2D}}26$ & \numprint{67108858} & \numprint{1149107290} & \numprint{45} & \cite{KaGen} \\
            & $\textsf{del}_{\textrm{3D}}26$ & \numprint{67108864} & \numprint{1042545824} & \numprint{40} & \cite{KaGen} \\
            & $\textsf{del}_{\textrm{2D}}26$ & \numprint{67108864} & \numprint{402653086} & \numprint{26} & \cite{KaGen} \\
            & $\textsf{rgg}_{\textrm{3D}}27$ & \numprint{134214672} & \numprint{1575628350} & \numprint{36} & \cite{KaGen} \\
            & $\textsf{rgg}_{\textrm{2D}}27$ & \numprint{134217728} & \numprint{2386714970} & \numprint{46} & \cite{KaGen} \\
            & $\textsf{del}_{\textrm{2D}}27$ & \numprint{134217728} & \numprint{606413354} & \numprint{14} & \cite{KaGen} \\
            & $\textsf{del}_{\textrm{3D}}27$ & \numprint{134217728} & \numprint{2085147648} & \numprint{40} & \cite{KaGen} \\
            & \textsf{kmerP1a} & \numprint{138896082} & \numprint{296930692} & \numprint{40} & \cite{FSMC} \\
            & \textsf{kmerA2a} & \numprint{170372459} & \numprint{359883478} & \numprint{40} & \cite{FSMC} \\
            & \textsf{kmerV1r} & \numprint{214004392} & \numprint{465409664} & \numprint{8} & \cite{FSMC} \\
        \midrule
            \multirow{15}{*}{\rotatebox[origin = c]{90}{High-degree graphs}}
            & \textsf{amazon} & \numprint{400727} & \numprint{4699738} & \numprint{2747} & \cite{SNAP} \\
            & \textsf{eu-2005} & \numprint{862664} & \numprint{32276936} & \numprint{68963} & \cite{DIMACS2012} \\ 
            & \textsf{youtube} & \numprint{1134890} & \numprint{5975246} & \numprint{28754} & \cite{SNAP} \\
            & \textsf{in-2004} & \numprint{1382867} & \numprint{27182946} & \numprint{21869} & \cite{DIMACS2012} \\
            & \textsf{com-orkut} & \numprint{3072441} & \numprint{234370166} & \numprint{33313} & \cite{SNAP} \\
            & \textsf{enwiki-2013} & \numprint{4203323} & \numprint{183879456} & \numprint{432260} & \cite{LWA} \\
            & \textsf{enwiki-2018} & \numprint{5608705} & \numprint{234488590} & \numprint{248444} & \cite{LWA} \\ 
            & \textsf{uk-2002} & \numprint{18483186} & \numprint{523574516} & \numprint{194955} & \cite{DIMACS2012} \\
            & \textsf{arabic-2005} & \numprint{22743881} & \numprint{1107806146} & \numprint{575628} & \cite{LWA} \\
            & \textsf{uk-2005} & \numprint{39454463} & \numprint{1566054250} & \numprint{1776858} & \cite{LWA} \\
            & \textsf{it-2004} & \numprint{41290648} & \numprint{2054949894} & \numprint{1326744} & \cite{LWA} \\
            & \textsf{twitter-2010} & \numprint{41652230} & \numprint{2405026092} & \numprint{2997487} & \cite{LWA} \\
            & \textsf{sk-2005} & \numprint{50636059} & \numprint{3620126660} & \numprint{8563816} & \cite{LWA} \\
            & \textsf{uk-2007-05} & \numprint{105153952} & \numprint{6603753128} & \numprint{975419} & \cite{DIMACS2012} \\
            & \textsf{webbase-2001} & \numprint{115554441} & \numprint{1709619522} & \numprint{816127} & \cite{LWA} \\
        \bottomrule
    \end{tabular} 
    \label{tab:graphs}
\end{table}

\newpage 
\section{Detailed Results for Large $k$}\label{s:exp:largek}

\begin{table*}[h]
    \centering
        \caption{Edge cut and running time results for $k \in \{2^{10}, 2^{14}, 2^{17}, 2^{20}\}$ and $\varepsilon = 3\%$ on Machine~A.
        The gmean imbalance of infeasible solutions is shown next to the number of infeasible solutions.
        The last two columns \emph{rel. time} and \emph{rel. cut} show the gmean running times and edge cuts relative to \algnameFast\ of all instances for which the respective algorithm does not crash (timeout instances are additionally excluded in edge cut comparisons).}
    \begin{tabular}{l|rrr|r|rr}
        Algorithm & \multicolumn{1}{c}{\# timeout} & \multicolumn{1}{c}{\# crash} & \multicolumn{1}{c}{\# infeasible}  & \multicolumn{1}{|c|}{\# feasible} & \multicolumn{1}{c}{rel. time} & \multicolumn{1}{c}{rel. cut} \\
        \midrule
            \algnameFast & \textbf{\numprint{\dkaminparFastTimeoutsLargek}} & \textbf{\numprint{\dkaminparFastFailsLargek}} & \textbf{\numprint{\dkaminparFastInfeasiblesLargek}} & \textbf{\numprint{\dkaminparFastFeasiblesLargek}} & 1.00 & 1.00 \\
        \midrule 
            \partitioner{ParHIP-Fast} & \numprint{\parhipFastTimeoutsLargek} & \numprint{\parhipFastFailsLargek} & \numprint{\parhipFastInfeasiblesLargek} (\parhipFastGmeanInfImbalanceLargek) & \numprint{\parhipFastFeasiblesLargek} & \parhipFastRunningTimeLargek & \parhipFastCutLargek \\
            \partitioner{ParHIP-Eco} & \numprint{\parhipEcoTimeoutsLargek} & \numprint{\parhipEcoFailsLargek} & \numprint{\parhipEcoInfeasiblesLargek} (\parhipEcoGmeanInfImbalanceLargek) & \numprint{\parhipEcoFeasiblesLargek} & \parhipEcoRunningTimeLargek & \parhipEcoCutLargek \\
            \partitioner{ParMETIS} & \textbf{\numprint{\parmetisTimeoutsLargek}} & \numprint{\parmetisFailsLargek} & \numprint{\parmetisInfeasiblesLargek} (\parmetisGmeanInfImbalanceLargek) & \numprint{\parmetisFeasiblesLargek} & \parmetisRunningTimeLargek & \parmetisCutLargek \\
        \midrule 
            \partitioner{KaMinPar} & \textbf{\numprint{\kaminparTimeoutsLargek}} & \textbf{\numprint{\kaminparFailsLargek}} & \textbf{\numprint{\kaminparInfeasiblesLargek}} & \textbf{\numprint{\kaminparFeasiblesLargek}} & \textbf{\kaminparRunningTimeLargek} & \kaminparCutLargek \\
        \bottomrule
    \end{tabular} 
    \label{tab:largek}
\end{table*}

As can be seen in \Cref{tab:largek}, only partitioners based on (distributed) deep MGP consistently compute feasible partitions with a large number of blocks.
\partitioner{ParHIP-Fast} and \partitioner{ParMETIS} only manage to do so on \R{\numprint{\parhipFastFeasiblesLargek}} and \R{\numprint{\parmetisFeasiblesLargek}} out of \R{\numprint{\dkaminparFastFeasiblesLargek}} instances, respectively.
We also note that \partitioner{ParHIP} becomes quite slow, with even its fast configuration being more than an order of magnitude slower than \algnameFast. 
This is because \partitioner{ParHIP} keeps a relatively larger number of vertices per block in the coarsest graph ($C = 5000$), and is therefore unable to shrink the graph sufficiently even for moderate values of $k$.

\newpage
\section{Edge Cut Results for Weak Scaling}\label{s:ws_cuts}

\begin{table*}[h]
     \caption{Edge cut results for weak scaling experiments on randomly generated graphs using \numprint{64}--\numprint{8192} cores of Machine~B and $\varepsilon = 3\%$.
     Some columns are not shown due to size constraints.
     Edge cuts and gmeans are reported relative to \algnameFast.
     Timeouts (15 minutes) and crashes are marked with \tblSymbTimeout\ and \tblSymbFailed, infeasible solutions are marked with \tblSymbInfeasible.
     }
     \centering
     \tiny
     \begin{tabular}{c|llrrrrr}
\multicolumn{1}{c}{} &&& \multicolumn{4}{c}{Cut on number of PEs / \numprint{1000}} & \\
\cmidrule{4-7}
\multicolumn{1}{c}{} & Graph & Algorithm & \numprint{128} & \numprint{512} & \numprint{2048} & \numprint{8192} & gmean \\
\midrule
\multirow{36}{*}{\rotatebox[origin = c]{90}{$k = 16$}} 
& \multirow{ 4 }{*}{ \rggTwoD{26}{8} } & \partitioner{dKaMinPar-Fast} & \numprint{280} & \numprint{572} & \numprint{1177} & \numprint{2428} & 1.00\\ 
& & \partitioner{ParHIP-Fast} & \numprint{1.10} & \numprint{1.13} & \numprint{1.14} & \tblSymbInfeasible & 1.11\\ 
& & \partitioner{ParMETIS} & \numprint{1.22} & \numprint{1.19} & \numprint{1.18} & \numprint{1.17} & 1.19\\ 
& & \partitioner{XtraPuLP} & \numprint{87.12} & \numprint{131.52} & \numprint{333.21} & \numprint{637.60} & 190.28\\ 
\cmidrule{2-8}
& \multirow{ 4 }{*}{ \rggTwoD{26}{32} } & \partitioner{dKaMinPar-Fast} & \numprint{4156} & \numprint{8544} & \numprint{17508} & \numprint{36050} & 1.00\\ 
& & \partitioner{ParHIP-Fast} & \numprint{1.12} & \numprint{1.14} & \numprint{1.16} & \tblSymbInfeasible & 1.15\\ 
& & \partitioner{ParMETIS} & \numprint{0.95} & \numprint{0.94} & \numprint{0.94} & \numprint{0.94} & 0.94\\ 
& & \partitioner{XtraPuLP} & \numprint{24.39} & \numprint{41.98} & \numprint{78.05} & \numprint{174.62} & 53.72\\ 
\cmidrule{2-8}
& \multirow{ 4 }{*}{ \rggThreeD{26}{8} } & \partitioner{dKaMinPar-Fast} & \numprint{4668} & \numprint{12037} & \numprint{31526} & \numprint{81200} & 1.00\\ 
& & \partitioner{ParHIP-Fast} & \numprint{1.01} & \numprint{1.01} & \numprint{1.00} & \tblSymbTimeout & 1.01\\ 
& & \partitioner{ParMETIS} & \numprint{0.95} & \numprint{0.96} & \numprint{0.92} & \numprint{0.92} & 0.94\\ 
& & \partitioner{XtraPuLP} & \numprint{12.48} & \numprint{19.92} & \numprint{25.31} & \numprint{37.47} & 21.13\\ 
\cmidrule{2-8}
& \multirow{ 4 }{*}{ \rggThreeD{26}{32} } & \partitioner{dKaMinPar-Fast} & \numprint{40411} & \numprint{103491} & \numprint{267885} & \numprint{701865} & 1.00\\ 
& & \partitioner{ParHIP-Fast} & \numprint{1.02} & \numprint{1.02} & \numprint{1.01} & \tblSymbInfeasible & 1.02\\ 
& & \partitioner{ParMETIS} & \numprint{0.88} & \numprint{0.87} & \numprint{0.86} & \numprint{0.84} & 0.87\\ 
& & \partitioner{XtraPuLP} & \numprint{6.25} & \numprint{10.03} & \numprint{11.41} & \numprint{15.71} & 9.94\\ 
\cmidrule{2-8}
& \multirow{ 4 }{*}{ \rhg{26}{8}{3.0} } & \partitioner{dKaMinPar-Fast} & \numprint{3} & \numprint{11} & \numprint{6} & \numprint{3} & 1.00\\ 
& & \partitioner{ParHIP-Fast} & \numprint{1.04} & \numprint{1.06} & \numprint{1.34} & \tblSymbInfeasible & 1.20\\ 
& & \partitioner{ParMETIS} & \numprint{4.85} & \tblSymbFailed & \tblSymbFailed & \tblSymbFailed & 5.51\\ 
& & \partitioner{XtraPuLP} & \numprint{9830.42} & \numprint{12480.89} & \numprint{77483.14} & \numprint{608259.52} & 48892.85\\ 
\cmidrule{2-8}
& \multirow{ 4 }{*}{ \rhg{26}{32}{3.0} } & \partitioner{dKaMinPar-Fast} & \numprint{76} & \numprint{66} & \numprint{70} & \numprint{90} & 1.00\\ 
& & \partitioner{ParHIP-Fast} & \numprint{1.06} & \numprint{1.13} & \numprint{1.18} & \tblSymbTimeout & 1.12\\ 
& & \partitioner{ParMETIS} & \numprint{4.22} & \numprint{6.11} & \numprint{8.95} & \tblSymbFailed & 6.07\\ 
& & \partitioner{XtraPuLP} & \numprint{1074.49} & \numprint{4613.94} & \numprint{17322.39} & \numprint{59519.97} & 6125.31\\ 
\midrule
\multirow{17}{*}{\rotatebox[origin = c]{90}{$2^{15}$ vertices per block}} 
& \multirow{ 4 }{*}{ \rggTwoD{26}{8} } & \partitioner{dKaMinPar-Fast} & \numprint{5543} & \numprint{22517} & \numprint{90764} & \numprint{364279} & 1.00\\ 
& & \partitioner{ParHIP-Fast} & \tblSymbInfeasible & \tblSymbInfeasible & \tblSymbInfeasible & \tblSymbFailed & --\\ 
& & \partitioner{ParMETIS} & \numprint{1.18} & \numprint{1.17} & \numprint{1.16} & \tblSymbFailed & 1.17\\ 
& & \partitioner{XtraPuLP} & \tblSymbInfeasible & \tblSymbInfeasible & \tblSymbFailed & \tblSymbFailed & --\\ 
\cmidrule{2-8}
& \multirow{ 4 }{*}{ \rggThreeD{26}{8} } & \partitioner{dKaMinPar-Fast} & \numprint{38348} & \numprint{156298} & \numprint{635175} & \numprint{2560517} & 1.00\\ 
& & \partitioner{ParHIP-Fast} & \numprint{1.03} & \numprint{1.03} & \tblSymbFailed & \tblSymbTimeout & 1.03\\ 
& & \partitioner{ParMETIS} & \numprint{0.99} & \numprint{1.00} & \numprint{0.99} & \tblSymbFailed & 1.00\\ 
& & \partitioner{XtraPuLP} & \tblSymbInfeasible & \tblSymbInfeasible & \tblSymbFailed & \tblSymbFailed & --\\ 
\cmidrule{2-8}
& \multirow{ 4 }{*}{ \rhg{26}{8}{3.0} } & \partitioner{dKaMinPar-Fast} & \numprint{1028} & \numprint{5157} & \numprint{19457} & \numprint{77084} & 1.00\\ 
& & \partitioner{ParHIP-Fast} & \tblSymbInfeasible & \tblSymbInfeasible & \tblSymbFailed & \tblSymbFailed & --\\ 
& & \partitioner{ParMETIS} & \numprint{2.45} & \tblSymbFailed & \tblSymbFailed & \tblSymbFailed & 2.52\\ 
& & \partitioner{XtraPuLP} & \numprint{117.45} & \tblSymbFailed & \tblSymbFailed & \tblSymbFailed & 115.92\\ 
\midrule
\multirow{17}{*}{\rotatebox[origin = c]{90}{$2^{12}$ vertices per block}} 
& \multirow{ 4 }{*}{ \rggTwoD{26}{8} } & \partitioner{dKaMinPar-Fast} & \numprint{16751} & \numprint{68202} & \numprint{277126} & \numprint{1109307} & 1.00\\ 
& & \partitioner{ParHIP-Fast} & \tblSymbInfeasible & \tblSymbFailed & \tblSymbFailed & \tblSymbFailed & --\\ 
& & \partitioner{ParMETIS} & \numprint{1.09} & \numprint{1.07} & \tblSymbFailed & \tblSymbFailed & 1.08\\ 
& & \partitioner{XtraPuLP} & \tblSymbFailed & \tblSymbFailed & \tblSymbFailed & \tblSymbFailed & --\\ 
\cmidrule{2-8}
& \multirow{ 4 }{*}{ \rggThreeD{26}{8} } & \partitioner{dKaMinPar-Fast} & \numprint{74560} & \numprint{300409} & \numprint{1213176} & \numprint{4876567} & 1.00\\ 
& & \partitioner{ParHIP-Fast} & \tblSymbInfeasible & \tblSymbFailed & \tblSymbFailed & \tblSymbFailed & 1.03\\ 
& & \partitioner{ParMETIS} & \numprint{1.00} & \tblSymbFailed & \tblSymbFailed & \tblSymbFailed & 1.01\\ 
& & \partitioner{XtraPuLP} & \tblSymbFailed & \tblSymbFailed & \tblSymbFailed & \tblSymbFailed & --\\ 
\cmidrule{2-8}
& \multirow{ 4 }{*}{ \rhg{26}{8}{3.0} } & \partitioner{dKaMinPar-Fast} & \numprint{7255} & \numprint{30901} & \numprint{123132} & \numprint{495273} & 1.00\\ 
& & \partitioner{ParHIP-Fast} & \tblSymbInfeasible & \tblSymbFailed & \tblSymbFailed & \tblSymbFailed & --\\ 
& & \partitioner{ParMETIS} & \numprint{1.67} & \tblSymbFailed & \tblSymbFailed & \tblSymbFailed & 1.67\\ 
& & \partitioner{XtraPuLP} & \tblSymbFailed & \tblSymbFailed & \tblSymbFailed & \tblSymbFailed & --\\ 
\bottomrule
\end{tabular} \\
     \label{tab:ws_cuts}
     \vspace*{-1.5cm} 
\end{table*}

\newpage 
\section{Edge Cut Results for Strong Scaling}\label{s:ss_cuts}

\begin{table*}[h]
        \caption{Edge cut results for strong scaling experiments on \numprint{64}--\numprint{8192} cores of Machine~B, $k = 16$, $\varepsilon = 3\%$. 
        Some columns are not shown due to size constraints.
        Edge cuts and gmeans are reported relative to \algnameFast.
        Timeouts (15 minutes) and crashes are marked with \tblSymbTimeout\ and \tblSymbFailed, infeasible solutions are marked with \tblSymbInfeasible.}
    \centering
    \footnotesize
    \begin{tabular}{llrrrrr}
&& \multicolumn{4}{c}{Cut on number of PEs / \numprint{1000}} & \\
\cmidrule{3-6}
Graph & Algorithm & \numprint{128}& \numprint{512}& \numprint{2048}& \numprint{8192}& gmean \\
\midrule
\multicolumn{7}{l}{High-degree graphs} \\
\midrule
\multirow{ 4 }{*}{ \textsf{webbase-2001} } & \partitioner{dKaMinPar-Fast} & \numprint{9634} & \numprint{9618} & \numprint{9602} & \numprint{9524} & 1.00\\ 
& \partitioner{ParHIP-Fast} & \numprint{1.09} & \numprint{1.10} & \numprint{1.14} & \numprint{1.12} & 1.11\\ 
& \partitioner{ParMETIS} & \tblSymbFailed & \tblSymbFailed & \tblSymbFailed & \tblSymbFailed & --\\ 
& \partitioner{XtraPuLP} & \numprint{2.62} & \numprint{3.30} & \numprint{6.13} & \numprint{13.79} & 5.20\\ 
\midrule
\multirow{ 4 }{*}{ \textsf{uk-2007-05} } & \partitioner{dKaMinPar-Fast} & \numprint{4093} & \numprint{4138} & \numprint{4176} & \numprint{4064} & 1.00\\ 
& \partitioner{ParHIP-Fast} & \numprint{1.07} & \numprint{1.03} & \numprint{1.01} & \numprint{1.06} & 1.04\\ 
& \partitioner{ParMETIS} & \tblSymbFailed & \tblSymbFailed & \tblSymbFailed & \tblSymbFailed & --\\ 
& \partitioner{XtraPuLP} & \numprint{42.94} & \numprint{116.57} & \numprint{185.19} & \numprint{264.43} & 125.13\\ 
\midrule
\multirow{ 4 }{*}{ \textsf{twitter-2010} } & \partitioner{dKaMinPar-Fast} & \numprint{616530} & \numprint{601073} & \numprint{604954} & \numprint{588380} & 1.00\\ 
& \partitioner{ParHIP-Fast} & \tblSymbTimeout & \tblSymbTimeout & \tblSymbTimeout & \tblSymbTimeout & --\\ 
& \partitioner{ParMETIS} & \tblSymbTimeout & \tblSymbFailed & \tblSymbFailed & \tblSymbFailed & --\\ 
& \partitioner{XtraPuLP} & \numprint{1.36} & \numprint{1.48} & \numprint{1.51} & \numprint{1.58} & 1.48\\ 
\midrule
\multicolumn{7}{l}{Low-degree graphs} \\
\midrule
\multirow{ 4 }{*}{ \textsf{kmer\_V1r} } & \partitioner{dKaMinPar-Fast} & \numprint{10936} & \numprint{10897} & \numprint{10880} & \numprint{10836} & 1.00\\ 
& \partitioner{ParHIP-Fast} & \tblSymbFailed & \tblSymbFailed & \tblSymbFailed & \tblSymbFailed & --\\ 
& \partitioner{ParMETIS} & \tblSymbFailed & \numprint{0.84} & \numprint{0.84} & \tblSymbTimeout & 0.84\\ 
& \partitioner{XtraPuLP} & \numprint{13.19} & \numprint{13.12} & \numprint{12.56} & \numprint{12.56} & 12.85\\ 
\midrule
\multirow{ 4 }{*}{ \textsf{nlpkkt240} } & \partitioner{dKaMinPar-Fast} & \numprint{5658} & \numprint{5641} & \numprint{5623} & \numprint{5547} & 1.00\\ 
& \partitioner{ParHIP-Fast} & \numprint{0.99} & \numprint{1.00} & \numprint{1.00} & \tblSymbFailed & 1.00\\ 
& \partitioner{ParMETIS} & \numprint{0.93} & \numprint{0.94} & \numprint{0.95} & \numprint{0.97} & 0.95\\ 
& \partitioner{XtraPuLP} & \numprint{3.44} & \numprint{3.88} & \numprint{2.80} & \numprint{2.43} & 3.09\\ 
\midrule
\multirow{ 4 }{*}{ $\textsf{rgg}_{\textrm{2D}}27$ } & \partitioner{dKaMinPar-Fast} & \numprint{349} & \numprint{349} & \numprint{350} & \numprint{347} & 1.00\\ 
& \partitioner{ParHIP-Fast} & \numprint{1.14} & \numprint{1.15} & \numprint{1.13} & \numprint{1.15} & 1.14\\ 
& \partitioner{ParMETIS} & \numprint{1.19} & \numprint{1.18} & \numprint{1.17} & \numprint{1.15} & 1.17\\ 
& \partitioner{XtraPuLP} & \numprint{72.87} & \numprint{150.46} & \numprint{259.72} & \numprint{194.77} & 153.46\\ 
\bottomrule
\end{tabular} 
    \label{tab:ss_cuts}
\end{table*}

\newpage
\section{Comparison against XtraPuLP}\label{s:xtrapulp}

We compare \algname\ against the single-level partitioner \partitioner{XtraPuLP}~\cite{XtraPuLP} (v0.3), which is a hybrid (OpenMPI+OpenMP) implementation of single-level label propagation.
To avoid excessive running time overheads due to graph generation, we only execute \partitioner{XtraPuLP} using a single thread per MPI process.

Strong scaling runnign times are shown in \Cref{fig:ss_xtrapulp}, while weak scaling runing times are shown in \Cref{fig:ws_xtrapulp_smallk} ($k = 16$) and \Cref{fig:ws_xtrapulp_largek} ($\{2^{15}, 2^{18}\}$ vertices per block).
The corresponding edge cuts are included in \Cref{tab:ss_cuts} (\Cref{s:ss_cuts}) and \Cref{tab:ws_cuts} (\Cref{s:ws_cuts}).
Surprisingly, \algname\ shows higher throughputs on most tested real-world and artificial graphs, although it should be noted that the performance of \partitioner{XtraPuLP} could likely be improved by running it with more threads per MPI process.
Moreover, we note that the edge cuts computed are often significantly worse (by up to 5 orders of magnitude on \rhg{3.0}{26}{8} partitioned on \numprint{8192} cores) than those of the multilevel partitioners.

\begin{figure*}[h]
    \centering 
    \input{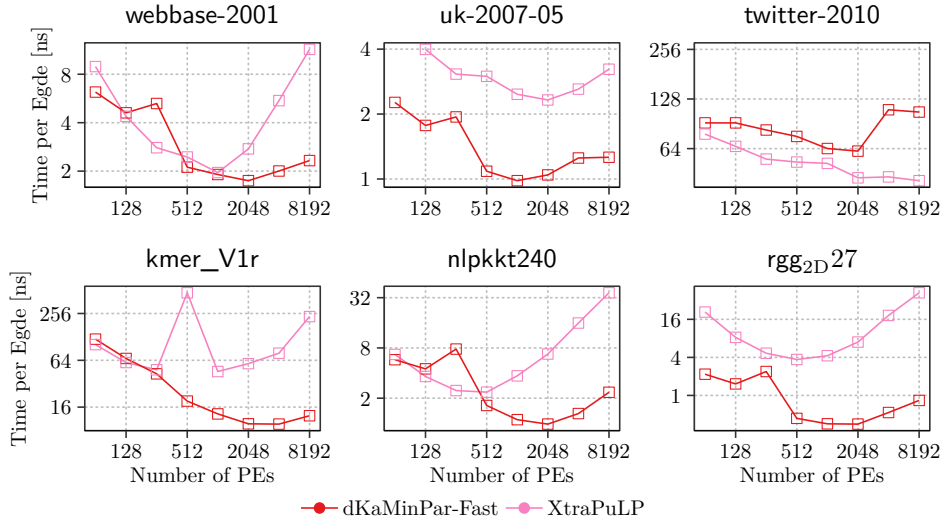}
\begin{tikzpicture}[x=1pt,y=1pt]
\definecolor{fillColor}{RGB}{255,255,255}
\begin{scope}
\definecolor{fillColor}{RGB}{255,255,255}

\path[fill=fillColor] (188.81,244.27) rectangle (206.15,261.62);
\end{scope}
\begin{scope}
\definecolor{drawColor}{RGB}{228,26,28}

\path[draw=drawColor,line width= 0.6pt,line join=round] (190.54,252.94) -- (204.42,252.94);
\end{scope}
\begin{scope}
\definecolor{drawColor}{RGB}{228,26,28}
\definecolor{fillColor}{RGB}{228,26,28}

\path[draw=drawColor,line width= 0.4pt,line join=round,line cap=round,fill=fillColor] (197.48,252.94) circle (  1.96);
\end{scope}
\begin{scope}
\definecolor{fillColor}{RGB}{255,255,255}

\path[fill=fillColor] (265.45,244.27) rectangle (282.80,261.62);
\end{scope}
\begin{scope}
\definecolor{drawColor}{RGB}{247,129,191}

\path[draw=drawColor,line width= 0.6pt,line join=round] (267.19,252.94) -- (281.06,252.94);
\end{scope}
\begin{scope}
\definecolor{drawColor}{RGB}{247,129,191}
\definecolor{fillColor}{RGB}{247,129,191}

\path[draw=drawColor,line width= 0.4pt,line join=round,line cap=round,fill=fillColor] (274.12,252.94) circle (  1.96);
\end{scope}
\begin{scope}
\definecolor{drawColor}{RGB}{0,0,0}

\node[text=drawColor,anchor=base west,inner sep=0pt, outer sep=0pt, scale=  0.80] at (206.15,250.19) {dKaMinPar-Fast};
\end{scope}
\begin{scope}
\definecolor{drawColor}{RGB}{0,0,0}

\node[text=drawColor,anchor=base west,inner sep=0pt, outer sep=0pt, scale=  0.80] at (282.80,250.19) {XtraPuLP};
\end{scope}
\end{tikzpicture}
    \caption{Strong scaling running times for the largest low- and high-degree graphs in our benchmark set, with $k = 16$, $\varepsilon = 3\%$ on \numprint{64}--\numprint{8192} cores of Machine~B.}
    \label{fig:ss_xtrapulp}
\end{figure*}

\begin{figure*}[h]
    \centering 
    \input{tex_plots/weakscaling_xtrapulp_smallk.tex}
\begin{tikzpicture}[x=1pt,y=1pt]
\definecolor{fillColor}{RGB}{255,255,255}
\begin{scope}
\definecolor{fillColor}{RGB}{255,255,255}

\path[fill=fillColor] (188.81,244.27) rectangle (206.15,261.62);
\end{scope}
\begin{scope}
\definecolor{drawColor}{RGB}{228,26,28}

\path[draw=drawColor,line width= 0.6pt,line join=round] (190.54,252.94) -- (204.42,252.94);
\end{scope}
\begin{scope}
\definecolor{drawColor}{RGB}{228,26,28}
\definecolor{fillColor}{RGB}{228,26,28}

\path[draw=drawColor,line width= 0.4pt,line join=round,line cap=round,fill=fillColor] (197.48,252.94) circle (  1.96);
\end{scope}
\begin{scope}
\definecolor{fillColor}{RGB}{255,255,255}

\path[fill=fillColor] (265.45,244.27) rectangle (282.80,261.62);
\end{scope}
\begin{scope}
\definecolor{drawColor}{RGB}{247,129,191}

\path[draw=drawColor,line width= 0.6pt,line join=round] (267.19,252.94) -- (281.06,252.94);
\end{scope}
\begin{scope}
\definecolor{drawColor}{RGB}{247,129,191}
\definecolor{fillColor}{RGB}{247,129,191}

\path[draw=drawColor,line width= 0.4pt,line join=round,line cap=round,fill=fillColor] (274.12,252.94) circle (  1.96);
\end{scope}
\begin{scope}
\definecolor{drawColor}{RGB}{0,0,0}

\node[text=drawColor,anchor=base west,inner sep=0pt, outer sep=0pt, scale=  0.80] at (206.15,250.19) {dKaMinPar-Fast};
\end{scope}
\begin{scope}
\definecolor{drawColor}{RGB}{0,0,0}

\node[text=drawColor,anchor=base west,inner sep=0pt, outer sep=0pt, scale=  0.80] at (282.80,250.19) {XtraPuLP};
\end{scope}
\end{tikzpicture}
    \caption{Throughput of $\textsf{rgg}_{\textrm{2D}}$, $\textsf{rgg}_{\textrm{3D}}$ and \textsf{rhg} graphs with $2^{26}$ vertices per compute node, average degree $\in \{8, 32\}$, $k = 16$ and $\varepsilon = 3\%$ on \numprint{64}--\numprint{8192} cores of Machine~B.}
    \label{fig:ws_xtrapulp_smallk}
\end{figure*}

\begin{figure*}[h]
    \centering 
    \input{tex_plots/weakscaling_xtrapulp_largek.tex}
\begin{tikzpicture}[x=1pt,y=1pt]
\definecolor{fillColor}{RGB}{255,255,255}
\begin{scope}
\definecolor{fillColor}{RGB}{255,255,255}

\path[fill=fillColor] (134.01,244.27) rectangle (151.35,261.62);
\end{scope}
\begin{scope}
\definecolor{drawColor}{RGB}{228,26,28}

\path[draw=drawColor,line width= 0.6pt,line join=round] (135.74,252.94) -- (149.62,252.94);
\end{scope}
\begin{scope}
\definecolor{drawColor}{RGB}{228,26,28}
\definecolor{fillColor}{RGB}{228,26,28}

\path[draw=drawColor,line width= 0.4pt,line join=round,line cap=round,fill=fillColor] (142.68,252.94) circle (  1.96);
\end{scope}
\begin{scope}
\definecolor{fillColor}{RGB}{255,255,255}

\path[fill=fillColor] (210.65,244.27) rectangle (228.00,261.62);
\end{scope}
\begin{scope}
\definecolor{drawColor}{RGB}{247,129,191}

\path[draw=drawColor,line width= 0.6pt,line join=round] (212.39,252.94) -- (226.26,252.94);
\end{scope}
\begin{scope}
\definecolor{drawColor}{RGB}{247,129,191}
\definecolor{fillColor}{RGB}{247,129,191}

\path[draw=drawColor,line width= 0.4pt,line join=round,line cap=round,fill=fillColor] (219.32,252.94) circle (  1.96);
\end{scope}
\begin{scope}
\definecolor{drawColor}{RGB}{0,0,0}

\node[text=drawColor,anchor=base west,inner sep=0pt, outer sep=0pt, scale=  0.80] at (151.35,250.19) {dKaMinPar-Fast};
\end{scope}
\begin{scope}
\definecolor{drawColor}{RGB}{0,0,0}

\node[text=drawColor,anchor=base west,inner sep=0pt, outer sep=0pt, scale=  0.80] at (228.00,250.19) {XtraPuLP};
\end{scope}
\begin{scope}
\definecolor{fillColor}{RGB}{255,255,255}

\path[fill=fillColor] (273.28,244.27) rectangle (290.63,261.62);
\end{scope}
\begin{scope}
\definecolor{drawColor}{RGB}{0,0,0}

\path[draw=drawColor,line width= 0.4pt,line join=round,line cap=round] (279.99,250.98) rectangle (283.92,254.91);
\end{scope}
\begin{scope}
\definecolor{fillColor}{RGB}{255,255,255}

\path[fill=fillColor] (320.57,244.27) rectangle (337.92,261.62);
\end{scope}
\begin{scope}
\definecolor{drawColor}{RGB}{0,0,0}

\path[draw=drawColor,line width= 0.4pt,line join=round,line cap=round] (329.25,256.00) --
	(331.89,251.42) --
	(326.60,251.42) --
	cycle;
\end{scope}
\begin{scope}
\definecolor{drawColor}{RGB}{0,0,0}

\node[text=drawColor,anchor=base west,inner sep=0pt, outer sep=0pt, scale=  0.80] at (290.63,250.19) {Feasible};
\end{scope}
\begin{scope}
\definecolor{drawColor}{RGB}{0,0,0}

\node[text=drawColor,anchor=base west,inner sep=0pt, outer sep=0pt, scale=  0.80] at (337.92,250.19) {Infeasible};
\end{scope}
\end{tikzpicture}
    \caption{Throughput of \textsf{rgg2D}, \textsf{rgg3D} and \textsf{rhg} graphs with $2^{26}$ vertices per compute node, average degree $8$, and $\varepsilon = 3\%$ on \numprint{64}--\numprint{8192} cores of Machine~B.
    The number of blocks are scaled with the size of the graph such that each block contains $2^{12}$ or $2^{15}$ vertices.}
    \label{fig:ws_xtrapulp_largek}
\end{figure*}

\fi
\end{document}